\documentclass[12pt,draftcls,onecolumn]{IEEEtran} 

\usepackage{times}
\usepackage{epsfig}
\usepackage{amsfonts}
\usepackage{amssymb}
\usepackage{amsbsy} 
\usepackage{amsmath}
\usepackage[ansinew]{inputenc}
\usepackage{pstricks}
\usepackage{lscape}
\usepackage{cite}
\usepackage{setspace}
\usepackage{rotating}
\usepackage{float}
\usepackage{booktabs,fixltx2e}
\usepackage{epstopdf}
\usepackage{upgreek}
\usepackage[font=small,skip=0pt]{caption}
\usepackage{xcolor}

\usepackage{hyperref}
\hypersetup{
    unicode=false,          
    pdftoolbar=true,        
    pdfmenubar=true,        
    pdffitwindow=false,     
    pdfstartview={FitH},    
    pdftitle={IEEE GRSM},    
    pdfauthor={Camps-Valls},   
    pdfsubject={GPR survey},   
    pdfcreator={Camps-Valls},  
    pdfproducer={Camps-Valls}, 
    pdfkeywords={keyword1} {key2} {key3}, 
    pdfnewwindow=true,      
    colorlinks=true,        
    linkcolor=blue,         
    citecolor=blue,        
    filecolor=blue,        
    urlcolor=blue          
}

\newcommand{\IG}{\includegraphics}
\newcommand{\x}{{\mathbf x}}

\newcommand{\f}{{\mathbf f}}
\newcommand{\bk}{{\mathbf k}}
\newcommand{\I}{{\mathbf I}}

\newcommand{\K}{{\mathbf K}}
\newcommand{\balpha}{\boldsymbol{\alpha}}

\definecolor{mycolor}{rgb}{0.1, 0.5, 0.2}

\begin{document}
 
\title{{Emulation as an Accurate Alternative to Interpolation in Sampling Radiative Transfer Codes}}

\author{Jorge Vicent, Jochem Verrelst, Juan Pablo Rivera-Caicedo, Neus Sabater, Jordi Mu\~noz-Mar\'i, 
Gustau Camps-Valls,~\IEEEmembership{IEEE Fellow} and Jos\'e Moreno,~\IEEEmembership{Senior Member,~IEEE}
\thanks{{\bf Preprint. Paper published in IEEE Journal of Selected Topics in Applied Earth Observations and Remote Sensing, vol. 11, no. 12, pp. 4918-4931, Dec. 2018, doi: 10.1109/JSTARS.2018.2875330.}. This project was carried out in the frame of ESA's project {\it FLEX L2 End-to-End Simulator Development and Mission Performance Assessment} ESA Contract No. 4000119707/17/NL/MP. Jochem Verrelst was supported by the European Research Council (ERC) under the ERC-2017-STG SENTIFLEX project (grant agreement 755617).  Jordi Mu\~noz-Mar\'i was support by MINECO/ERDF under Grant CICYT TIN2015-64210-R. Gustau Camps-Valls was supported by the ERC under the ERC-CoG-2014 SEDAL project (grant agreement 647423).}

\thanks{}
\thanks{J. Vicent, J. Verrelst, N. Sabater, J. Mu\~noz-Mar\'i, G. Camps-Valls, and J. Moreno are with Image Processing Laboratory, University of Valencia, 46980 Paterna (Valencia), Spain. Web: http://ipl.uv.es. E-mail: \{jorge.vicent, jochem.verrelst, m.neus.sabater, jordi.munoz, gustau.camps, jose.moreno\}@uv.es.}
\thanks{J.P. Rivera is with CONACYT-UAN, Departamento: Secretaria de investigaci\'on y posgrado, Tepic, Nayarit, Mexico. Email: jprivera@conacyt.mx}
}

\markboth{IEEE Journal of Selected Topics in Applied Earth Observations and Remote Sensing, Vol. XX, No. Y, Month YYYY}
{Vicent et al.: Emulation as an accurate alternative *to* interpolation in sampling
radiative transfer codes}

\maketitle

\begin{abstract}
{Computationally expensive Radiative Transfer Models (RTMs) are widely used} to realistically reproduce the light interaction with the Earth surface and atmosphere. Because these models take long processing time, the common practice is to first generate a sparse look-up table (LUT) and then make use of interpolation methods to sample the multi-dimensional LUT input variable space. However, the question arise whether common interpolation methods perform most accurate. 
As an alternative to interpolation, this work proposes to use emulation, i.e., approximating the RTM output by means of statistical learning. Two experiments were conducted to assess the accuracy in delivering spectral outputs using interpolation and emulation: (1) at canopy level, using PROSAIL; and (2) at top-of-atmosphere level, using MODTRAN. Various interpolation (nearest-neighbour, inverse distance weighting, piece-wice linear) and emulation (Gaussian process regression (GPR), kernel ridge regression, neural networks) methods were evaluated against a dense reference LUT.
In all experiments, the emulation methods clearly produced more accurate output spectra than classical interpolation methods. GPR emulation performed up to ten times more accurately than the best performing interpolation method, and this with a speed that is competitive with the faster interpolation methods. 
It is concluded that emulation can function as a fast and more accurate alternative to commonly used interpolation methods for reconstructing RTM spectral data. 
\end{abstract}



\section{Introduction}\label{sec:intro}
 
{Physically-based radiative transfer models (RTMs) allow remote sensing scientists to understand the light interactions between water, vegetation and atmosphere \cite{hydrolight,Zhang2004,Jacquemoud2009}. RTMs are physically-based computer models that describe scattering, absorption and emission processes in the {visible to microwave} region \cite{Deiveegan2008,VanDerTol2014}. These models are widely used in applications such as (i) inversion of atmospheric and vegetation properties from remotely sensed data (see \cite{Verrelst2015b} for a review), (ii) to generate artificial scenes as would be observed by a sensor \cite{Verhoef2012,Meharrar2014,Vicent2016}, and (iii) sensitivity analysis  of RTMs \cite{Verrelst2015a}.} 
In the optical domain, a diversity of vegetation, atmosphere and water RTMs have continuous been improved in accuracy from simple semi-empirical RTMs towards advanced ray tracing RTMs. 
This evolution has {led to an increase in} complexity, intepretability and computational requirements to run the model, which bears implications towards practical applications. On the one hand, computationally cheap RTMs are models with relatively few input parameters that enables fast calculations (e.g., \cite{Feret08,Verhoef1984} for vegetation and \cite{Rahman1994,Lipton2009} for atmosphere). 
On the other hand, computationally expensive RTMs are complex physically-based mathematical models with a large number of input variables. In short, the following families of RTMs can be considered as computationally expensive: 
(1) Monte Carlo ray tracing models (e.g., Raytran \cite{Govaerts1998}, FLIGHT \cite{North1996} and librat \cite{Lewis1999}),
 (2) voxel-based models (e.g., DART \cite{Gastellu1996})
and (3) advanced integrated vegetation and atmospheric transfer models that consists of various subroutines (e.g., SimSphere \cite{Petropoulos2009}, SCOPE \cite{VanDerTol2009} and MODTRAN \cite{Berk2006}).
Despite the higher accuracy of these RTMs  to model the light-vegetation and atmosphere interactions (see e.g., \cite{Govaerts1998,Espana1999}), their high computational burden make them impractical for {practical applications that demand many simulations}, and alternatives have to be sought.
 
In order to overcome this limitation, RTMs are most commonly applied by means of look-up tables (LUTs)~\cite{Verrelst2015b}. LUTs are pre-stored RTM output data so that the computational of the RTM has to be done only one time, prior to the application. Nevertheless, for reasons of memory storage and processing time, LUTs are usually kept to a reasonable size, especially in case of computationally expensive RTMs. The common approach is then to seek through the multi-dimensional LUT input variable space by means of interpolation techniques. Various interpolation techniques have been developed both for gridded and scattered datasets~{\cite{interpolationref,amidror,Ientilucci2008}}. Linear interpolation is the most used approach in both gridded and scattered datasets due to its balance between processing speed and accuracy~\cite{guanter2009,Lyapustin2011,Scheck2016}. However, the main drawback of linear interpolation in high dimensional scattered datasets is that the underlying triangulation is computationally expensive and uses large computer memory. 

Emulation of costly codes is an alternative to interpolation, but based on statistical principles. The core idea of an emulator is to extract (or \textit{learn}) the statistical information from a limited set of simulations of the original deterministic model \cite{OHagan2006,Gomez-Dans2015}. Emulators then approximate the original RTM at a tiny fraction of its speed and this {can be readily applied} in tedious processing routines \cite{Verrelst2016GSA,Verrelst2017}. The use of emulators deals with some extra advantages such as the use of a non-gridded input parameter space, making it  more versatile than several interpolation methods (e.g., piece-wise cubic splines).
In different research fields, such as engineering, energy, robotics, and environmental modeling, emulators have already demonstrated to be a more efficient alternative to classical LUT interpolation methods~\cite{Busby09,Kim16energybuildings,Razavi12,Bastos09,OHagan2006,OHagan2012,Owen17}.
However, these studies are generally limited to models with a few output dimensions, low levels of noise, or low degrees of freedom and ill-posedness. Therefore, an important question arises here whether emulators are able to compete with interpolation methods in sampling capabilities of hyperspectral RTM outputs, both in terms of accuracy and processing speed. The problem is thus new and actually challenges the potential capabilities of emulators. This brings us to the main objective of this work, i.e., to analyze the performance of emulators as an alternative of classical RTM-based LUT interpolation for {sampling the LUT parameter space}.
To do so, two contrasting LUT spectral outputs are examined: one of spectrally smooth top-of-canopy (TOC) reflectance data, and another of more sharper (high frequency bands) top-of-atmosphere (TOA) radiance data. We give experimental evidence that emulation generally outperforms interpolation in both computational cost and accuracy, which suggest they might be better suited for RTM-based LUT sampling.

The remainder of this paper is as follows. Section \ref{sec:AL} gives a theoretical overview of the analyzed interpolation and emulation methods. Section \ref{sec:MaterialsMethods} presents the materials and methods to study the performance of interpolation and emulation methods in terms of accuracy and computation time. This is followed by presenting the results in Section \ref{sec:results} which are discussed in a broader context (Section \ref{sec:discussion}). Section \ref{sec:conclude} concludes this paper.

\section{Interpolation \& emulation theory}\label{sec:AL}
In this section, we first present common interpolation methods (Section \ref{interpolation}), and then address the emulation theory (Section \ref{emulation}). 

\subsection{Interpolation}\label{interpolation}
Let us consider a $D$-dimensional input space $\mathcal{X}$ from where we sample $\x \in\mathcal{X}\subset \mathbb{R}^D$ in which a $K$-dimensional object function $\f(\x;\lambda) = [f(\x;\lambda_1),\ldots,f(\x;\lambda_K)]: \mathbb{R}\mapsto \mathbb{R}^K$ is evaluated. In the context of this paper, $\mathcal{X}$ comprises the $D$ input variables (e.g., Leaf Area Index (LAI), Aerosol Optical Thickness (AOT), Visual Zenith Angle (VZA)) that control the behavior of the function $\f(\x;\lambda)$, i.e., a water, canopy or atmospheric RTM. Here, $\lambda$ represents the wavelengths in the $K$-dimensional output space\footnote{For sake of simplicity, the wavelength dependency is omitted in the formulation in this paper, i.e., $\f(\x;\lambda)\equiv\f(\x)$.}.
An interpolation, $\widehat\f(\x)$, is therefore a technique used to approximate model simulations, $\f(\x)=\widehat\f(\x)+\varepsilon$, based on the numerical analysis of an existing set of {\it nodes}, $\f_i=\f(\x_i)$, conforming a pre-computed LUT. The concept of interpolation has been widely used in remote sensing applications, including retrieval of biophysical parameters \cite{Gastellu2003} and atmospheric correction algorithms \cite{guanter2009,Lyapustin2011}. The following non-exhaustive list gives an overview of commonly used interpolation techniques in remote sensing:

\begin{itemize}
    \item {\bf Nearest-neighbor:} This is the simplest method for interpolation, which is based on finding the closest LUT {\it node} $\x_i$ to a query point $\x_q$ (e.g., by minimizing their Euclidean distance) and associate their output variables, i.e., $\widehat\f(\x_q)=\f(\x_i)$. This fast method is valid for both gridded and scattered LUTs.  
    \item {\bf Inverse Distance Weighting (IDW)} \cite{shepard}{\bf:} Also known as {\it Shepard's method}, this method weights the $n$ closest LUT nodes to the query point $\x_q$ (see equation \ref{eqn:inverseinterp1}) by the inverse of the distance metric $d(\x_q,\x_i): \mathcal{X}\mapsto \mathbb{R}^+$ (e.g., the Euclidean distance):
    
    \begin{equation}
    \label{eqn:inverseinterp1}
    \widehat\f(\x_q) = \frac{\sum_{i=1}^n\omega_i\f(\x_i)}{\sum_{i=1}^n\omega_i} ,
    \end{equation}
    
    where $\omega_i = d(\x_q,\x_i)^{-p}$, and $p$ (typically $p$=2) is a tuneable parameter known as {\it power parameter}. {When $p$ is large, this method produces the same results as the nearest-neighbor interpolation}. The method is computationally cheap but it is affected by LUT nodes far from the query point. The modified Shepard's method \cite{lukaszyk2004} aims to reduce the effect of distant {\it grid points} by modifying the weights by equation \ref{eqn:inverseinterp3}:
    
    \begin{equation}
    \label{eqn:inverseinterp3}
    \omega_i = \left(\frac{R-d(\x_q,\x_i)}{R\cdot d(\x_q,\x_i)}\right)^p ,
    \end{equation}
    
    where $R$ is the maximum Euclidean distance to the $n$ closest LUT nodes.
    \item {\bf Piece-wise linear:}  This method is commonly used in remote sensing applications due to its balance between computation time and interpolation error ~\cite{Cooley20021414,Richter2002,guanter2009}.
The implementation of the  linear interpolation is based on the Quickhull algorithm~\cite{Barber1996469} for triangulations in multi-dimensional input spaces. 
For the scattered LUT input data, the piece-wise linear interpolation method is reduced to find the corresponding Delaunay simplex~\cite{Delaunay1934} (e.g., a triangle when $D=2$) that encloses a query $D$-dimensional point $\x_q$ (see equation \ref{eqn:lininterp} and Fig. \ref{fig:delaunay}):

\begin{equation} \label{eqn:lininterp}
\widehat{\f}_i(\x_q) = \sum_{j=1}^{D+1}\omega_j\f(\x_j) ,
\end{equation}

where $\omega_j$ are barycentric coordinates of $\x_q$ with respect to the $D$-dimensional simplex (with $D+1$ vertices) \cite{coxeter}. {Notice that, though similar, IDW with parameters $n$=$D+1$ and $p$=1 is not strictly the same as linear interpolation since IDW uses Euclidean distances instead of the barycentric coordinates in linear interpolation.}
Since $\f(\x)$ is a $K$-dimensional function, the result of the interpolation is also $K$-dimensional.

\begin{figure}[!ht]
	\centering
	\IG[width=\linewidth]{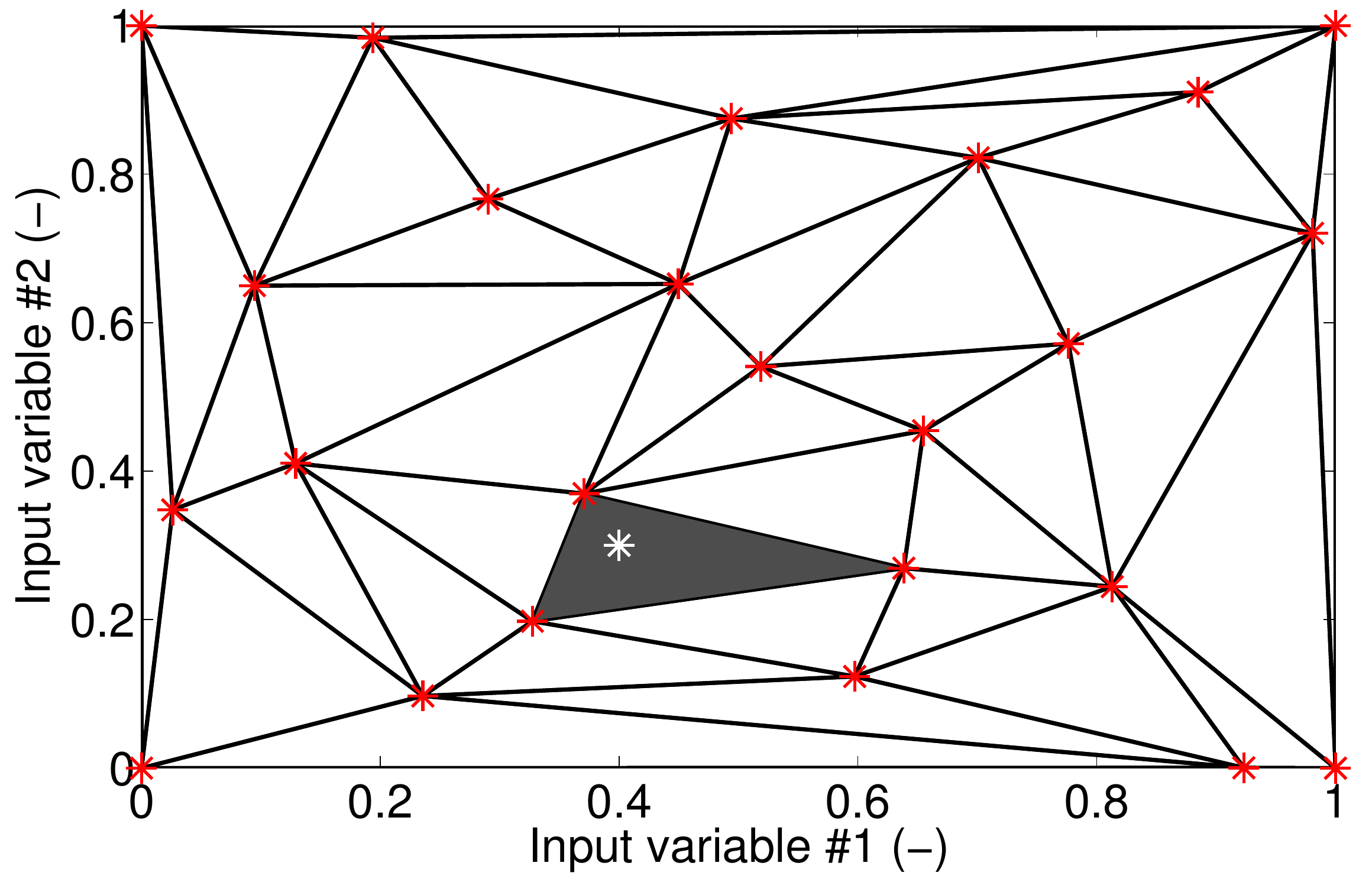}
	\caption{Schematic representation of a $2$-dimensional interpolation of a query point $\x_q$ (white $\ast$) after Delaunay triangulation (solid lines) of the scattered LUT nodes $X_i$ (\red{$\ast$}).}
	\label{fig:delaunay}
\end{figure}
   
    In scattered LUTs, the underlying Delaunay triangulation is computationally expensive in high dimensional input spaces (typically $D>$6) and is also limited by its intensive memory consumption \cite{Barber1996469,matlabgriddatan}.
\end{itemize}

Other existing advanced interpolation methods (e.g., Sibson's interpolation \cite{sibson,Park2006} and piece-wise cubic splines~\cite{Bartels}) were not considered in the analysis due to their even more intensive memory consumption in high dimensional input spaces. 

\subsection{Emulation}\label{emulation}

Emulation is a statistical learning technique used to estimate model simulations when the model under investigation is too computationally costly to be run many times \cite{OHagan2006}.  
The concept of developing emulators have already been applied in the last few decades in the climate and environmental modeling communities \cite{Petropoulos2009,Rohmer2011,carnevale2012,villa2012,castelletti2012,lee2013,bounceur2014,Ireland2015}.
The basic idea is that an emulator uses a limited number of simulator runs, i.e., input-output pairs (corresponding to training samples), {to train a machine learning regression algorithm in order} to infer the values of the complex simulator output given a yet-unseen input configuration. These training data pairs should ideally cover the multidimensional input parameter space using a space-filling sampling algorithm, e.g., Latin hypercube sampling~\cite{McKay1979}. 

As with LUT interpolation, once the emulator is built, it is not necessary to perform any additional runs of the model; the emulator computes the output that is otherwise generated by the simulator \cite{OHagan2006}. Accordingly, emulators are statistical models that can generalize the input-output relations from a subset of examples generated by RTMs to unseen data. Note that building an emulator is essentially nothing more than building an advanced regression model as typically done for biophysical parameter retrieval applications (see pioneering works using neural networks (NNs) \cite{Baret1995,Combal2003} and also more recent statistical methods \cite{Verrelst12a,Rivera14,Verrelst2015b}, but in reversed order: whereas a retrieval model converts input spectral data (e.g., reflectance) into one or more output biophysical variables, an emulator converts input biophysical variables into output spectral data.
When it comes to emulating RTM spectral outputs, however, the challenge lies in delivering a full spectrum, i.e., predicting {contiguous} spectral bands. This is an additional difficulty compared to traditional interpolation methods or standard emulators that only deliver one output~\cite{hankin2005}. It bears the consequence that the machine learning methods should be able to generate multiple outputs to be able reconstructing a full spectral profile. This is not a trivial task. For instance, the full, contiguous spectral profile between 400 an 2500 nm consists of over 2000 bands when binned to 1~nm resolution.
 
Not all regression models are able to deal with high dimensional outputs. Only some of them can obtain multi-output models. For instance, with NNs it is possible to train multi-output models. However, training a complex multi-output statistical model with the capability to generate so many output bands would take considerable computational time and would probably incur in a certain risk of overfitting because of model over-representation. A workaround solution has to be developed that enables the regression algorithms to cope with large, spectroscopy datasets. An efficient solution is to take advantage of the so-called \emph{curse of spectral redundancy}, i.e., the Hughes phenomenon~\cite{Hughes1968}. Since spectroscopy data typically shows a great deal of collinearity, it implies that such data can be converted to a lower-dimensional space through dimensionality reduction (DR) techniques. Accordingly, spectroscopy data can be converted into components, which are only a fraction of the original amount of bands, and implies that the multi-output problem is greatly reduced to a number of components that preserve the spectral information content. Afterwards the components {are then again reconstructed to spectral data. 
In this paper, we first apply a Principal Component Analysis (PCA)~\cite{Wold87} to the spectral data in order to reduce it to a given number of features (components). This step greatly reduces the number of dimensions while keeping 99\% of the spectra variance. Through dimensionality reduction, the problem is better conditioned and allows to either train multi- or single-output models on this reduced set of components~\cite{bounceur2014,Rivera2015,Gomez-Dans2015,Verrelst2016GSA,Verrelst2017}. As the models are trained to predict on the reduced set of components, the final step of the process is to project back the predictions to the original spectra space by applying the inverse PCA}.

\subsection{Machine learning regression algorithms}

Two steps are required to enable approximating a RTM through emulation. The first step involves building a statistically-based representation (i.e., an emulator) of the RTM using statistical learning from a set of training data points derived from runs of the actual model under study (LUT nodes in the context of interpolation).  
The second step uses the emulator previously built in the first step to compute the output in the LUT input parameter space that would otherwise have  
to be generated by the original RTM \cite{OHagan2006}.
Based on the literature review above and earlier emulation evaluation studies \cite{Rivera2015,Verrelst2016GSA,Verrelst2017}, the following three machine learning methods potentially serve as powerful methods to function as accurate emulators, being: (1) Gaussian processes regression (GPR), (2) kernel ridge regression (KRR) and (3) neural networks (NNs).

We selected these three methods as representative
{state-of-the-art machine learning families for regression. Kernel ridge regression generalizes linear regression via kernel functions. Gaussian Process Regression is essentially the probabilistic version of KRR, and has been widely used for biogeophysical parameter and emulation \cite{Bastos09,Conti2009,Liu2009}.}
Neural networks are standard approximation tools in statistics and artificial intelligence, and are currently revived through the popular adoption of deep learning models~\cite{deepl}. We explore all these techniques for the sake of a complete benchmarking of standard methods available. These methods are briefly outlined below.

{
Kernel methods in machine learning owe their name to the use of kernel functions~\cite{ShaweTaylor04,CampsValls09wiley,Rojo17dspkm}. These functions quantify similarities between input samples of a dataset. Similarity reproduces a linear dot product (scalar) computed in a possibly higher dimensional feature space, yet without ever computing the data location in the feature space. The following two methods are gaining increasing attention: (1) GPR generalize Gaussian probability distributions in function spaces~\cite{Rasmussen2006}, and (2) KRR, which perform least squares regression in feature spaces~\cite{Suykens1999}.
The expressions defining the weights and the predictions obtained by GPR and KRR are the same, but interestingly these expressions are obtained following different approaches. GPR follow a probabilistic approach (see~\cite{Rasmussen2006}), whereas KRR implement a discriminative approach for regression and function approximation. In both cases, the prediction and the predictive variance of the model for new samples are given by
\begin{align} \label{eqn:gpr}
    \widehat\f(\x_q) &= \sum_{i=1}^n\balpha_i k(\x_i,\x_q) \\
    \mathbb{V}[\widehat\f(\x_q)] &= k(\x_q,\x_q) - \bk_*^\top(K+\sigma_n^2\I)^{-1}\bk_*
\end{align}
where $k(\cdot,\cdot)$ is a covariance (or kernel function), $\bk_*$ is the vector of covariances between the query point, $\x_q$, and the $n$ or training points, and $\sigma_n^2$ accounts for the noise in the training samples. As one can see, the prediction is obtained as a linear combination of weighted kernel (covariance) functions, the optimal weights given by $\balpha=(\K+\sigma_n^2\I)^{-1}\f(\x)$.
Many different functions can be used as kernels for both GPR and KRR. In this paper we used a standard Gaussian radial basis function (RBF) kernel for KRR, which has a single length hyperparameter for all input dimensions, and the automatic relevance determination squared exponential (ARD-SE) kernel for GPR, which has a separate length hyperparameter for each input dimension. For KRR, these hyperparameters are tuned through standard cross-validation techniques are used to choose the best hyperparameters. For GPR, stochastic gradient descent algorithms maximizing the marginal log-likelihood are employed, which allow to optimize a large number of hyperparemeters (compared to KRR) in a computational effective way.
}

NNs are essentially fully connected layered structures of artificial neurons~\cite{Haykin1999}. A NN is a (potentially fully) connected structure of neurons organized in layers. Neurons of different layers are interconnected with the corresponding links (weights).
{
The output on the final layer of the NN, and thus the prediction, is given by
\begin{equation} \label{eqn:nn}
    \widehat\f(\x_*) = g\Bigg( \sum_{k=1}^n w_{jk}^l x_k^{l-1} + b_k^l \Bigg)
\end{equation}
where $w_{jk}^l$ and $b_k^l$ are the weights and bias at the $l^{th}$ layer, respectively, $x_k^{l-1}$ is the input vector at the $l-1^{th}$ layer, and $g$ is an activation function, which at the output layer and for regression problems could be the identity function}. Training a NN implies selecting a structure (number of hidden layers and nodes {\em per} layer), initialize the weights, shape of the nonlinear activation function, learning rate, and regularization parameters to prevent overfitting~\cite{Bishop95}. The selection of a training algorithm and the loss function both have an impact on the final model. In this work, we  used the standard multi-layer perceptron, which is a fully-connected network. We selected just one hidden layer of neurons. We optimized the NN structure using the Levenberg-Marquardt learning algorithm with a squared loss function.

{{One can note the similarity between the prediction functions used in the emulators (see equations \ref{eqn:gpr} and \ref{eqn:nn}) with those used for interpolation (see section \ref{interpolation}).
In fact, the theoretical relation between both approaches was extensively discussed back in 1970 in the context of splines and Gaussian processes regression in ~\cite{kimeldorf1970}. Essentially, machine learning emulators perform regression and hence are more flexible functions for fitting than interpolation, as the solution is not forced to pass through the observed points. On the downside, the emulation approach may be hampered by an adequate estimation of the hyperparameters (i.e. regularization). When a good estimate of the hyperparameters is achieved emulation should obtain equal or better results than interpolation, otherwise the regression may incur in a certain risk of overfitting.}

\section{Materials and Methods}\label{sec:MaterialsMethods}
In this section we will start by giving an overview the software used to generate the synthetic datasets used to assess the performance of the interpolation/emulation methods (Sections \ref{ARTMO} and \ref{RTMsim}). We will continue by describing these datasets (Section \ref{analysis}) and finish by explaining the error metrics used to evaluate the performance of the interpolation/emulation methods (Section \ref{validation}).

\subsection{The ARTMO and ALG toolboxes}\label{ARTMO}

This study was conducted within two in-house developed graphical user interface (GUI) software packages named ARTMO (Automated Radiative Transfer Models Operator)~\cite{Verrelst12c} and ALG (Atmospheric Look-up table Generator)~\cite{Vicent2017}. 
Both software packages facilitate the usage of a suite of leaf, canopy and atmosphere radiative transfer models (RTMs) including, among others, PROSAIL (i.e., the leaf model PROSPECT coupled with the canopy model SAIL~\cite{Jacquemoud09}) and MODTRAN5. As a novelty, the latest ARTMO version (v. 3.24) is coupled with ALG (v. 1.2), which allows generating large multi-dimensional LUTs of TOA radiance data for Lambertian surfaces.

ARTMO also embodies a set of retrieval toolboxes, and recently an 'Emulator toolbox' was added~\cite{Rivera2015}. In the Emulator toolbox, several of those MLRAs can be trained by RTM-generated LUTs, whereby biophysical variables are used as input in the regression model, and spectral data is generated as output. In addition, ALG includes a function that allows various methods of interpolating gridded and scattered LUTs (i.e., nearest neighbor, piece-wise linear/splines, IDW).
The ARTMO and ALG packages are developed in MATLAB and can be freely downloaded from~\url{http://ipl.uv.es/artmo/}.

\subsection{Description of Simulated Datasets}\label{RTMsim}

\subsubsection{PROSPECT-4}

The leaf optical model PROSPECT-4~\cite{Feret08} calculates leaf reflectance and transmittance as a function of four biochemistry and anatomical variables: leaf structure ($N$), equivalent water thickness ($Cw$), chlorophyll content ($Cab$) and dry matter content ($Cm$). PROSPECT-4 simulates directional reflectance and transmittance over the solar spectrum from 400 to 2500 nm at the fine spectral resolution of 1~nm. 

\subsubsection{SAIL}

At the canopy scale, SAIL~\cite{Verhoef1984} approximates the RT equation through two direct fluxes (incident solar flux and radiance in the viewing direction) and two diffuse fluxes (upward and downward hemispherical flux)~\cite{verhoef2007}. SAIL input variables consist of leaf area index ($LAI$), leaf angle distribution ($LAD$), ratio of diffuse and direct radiation ($skyl$), soil coefficient (\emph{soil coeff.}), hot spot and sun-target-sensor geometry, i.e., solar/view zenith angle and relative azimuth angle (\emph{SZA}, \emph{VZA} and \emph{RAA}, respectively). Spectral input consists of leaf reflectance and transmittance spectra and a soil reflectance spectrum. The leaf optical properties can come from a leaf RTM such as PROSPECT, which results in the leaf-canopy model PROSAIL~\cite{Jacquemoud2009}. PROSAIL allows analyzing the impact of leaf biochemical variables on the hemispherical and bidirectional TOC reflectance.

\subsubsection{MODTRAN5}

At the atmosphere scale, MODTRAN5~\cite{Berk2006}, the moderate resolution transmittance code, is one of the most widely used radiative transfer codes in the atmospheric community due to its accurate simulation of the coupled absorption and scattering effects\cite{Stamnes:88,Goody1989539}. 
MODTRAN solves the RT equation in a multilayered spherically symmetric atmosphere by including the effects of molecular and particulate absorption/emission and scattering, surface reflections and emission, solar/lunar illumination, and spherical refraction.

\subsection{Experimental setup}\label{analysis}

Here we outline the experimental setup for running the interpolation and emulation experiments. For both PROSAIL and MODTRAN RTMs, LUTs were generated by means of Latin hypercupe sampling (LHS) within the RTM variable space with minimum and maximum boundaries as given in Tables~\ref{table_inputs} and \ref{table_inputs2}. {The selected input variables were chosen given their influence in both the entire spectra (e.g., Aerosol Optical Thickness, \r{A}ngstr\"om exponent) and in specific absorption bands (e.g., Chlorophyll absorption, water vapour, ozone).} A LHS of training data is preferred, as LHS covers the full parameter space, and thus, in principle, assures that the developed emulator/interpolation will be able to reconstruct correct spectral output for any possible combination of input variables. For both the canopy and atmospheric RTMs, three sizes of LUTs were created given the same LUT boundaries: 500, 2000 and 5000. While the most dense LUT (5000) was used as a reference LUT to evaluate the performances of the emulation and interpolation algorithms, the first two LUTS (500 and 2000) where simulated to actually run the emulation and interpolation techniques for different LUTs sizes. Additionally, the 64 vertex of the input variable space (i.e., where the input variables get the minimum/maximum values) were added to these two LUTs. The addition of these vertex enables consistent functioning of all tested interpolation techniques, i.e., that the input variable space is bounded and no extrapolation is performed.

\begin{table}[ht!] 
\begin{center}\small
    \caption{Range of vegetation input variables for the PROSAIL LUTs according to Latin Hypercube sampling. SAIL fixed variables: hot spot: 0.01; solar zenith angle: 30$^{\circ}$; observer zenith angle: 0$^{\circ}$; azimuth angle: 0$^{\circ}$.}
    \resizebox{\linewidth}{!}{ 
     \begin{tabular}{llccc@{}}
\hline
       \multicolumn{2}{c}{\bf Model variables} & {\bf Units} & {\bf Minimum} & {\bf Maximum}\\
\hline			
\multicolumn{5}{l}{\bf Leaf variables (PROSPECT-4)} \\
N &  Leaf structure index & unitless & 1.3 & 2.5 \\
Cw &  Leaf water content & [cm] & 0.002 & 0.05 \\
Cab & Leaf chlorophyll content & [${\mu}$g/cm$^{2}$]  & 1 & 70 \\
Cm &  Leaf dry matter content & [g/cm$^{2}$] & 0.002 & 0.05 \\
\\
\multicolumn{5}{l}{\bf Canopy variables (SAIL)}\\
LAI & Leaf area index & [m$^{2}$/m$^{2}$] & 0.1 & 7 \\
LAD & Leaf angle distribution & [$^{\circ}$] & 0 & 90 \\
\hline
	   \end{tabular}}
	   \label{table_inputs}
\end{center}
\end{table}

\begin{table}[ht!] 
\begin{center}\small
    \caption{Range of atmospheric input variables for the MODTRAN LUTs according to Latin Hypercube sampling. MODTRAN fixed geometric variables: solar zenith angle: 55$^{\circ}$; observer zenith angle: 0$^{\circ}$; azimuth angle: 0$^{\circ}$. {Remaining MODTRAN parameters were set to their default values.}}
    \resizebox{\linewidth}{!}{ 
     \begin{tabular}{llccc@{}}
\hline
       \multicolumn{2}{c}{\bf Model variables} & {\bf Units} & {\bf Minimum} & {\bf Maximum}\\
\hline			
O3C &  O$_3$ column concentration & [amt-cm] & 0.2 & 0.45 \\
CWV &  Columnar Water Vapour & scale-factor & 1 & 4 \\
AOT &  Aerosol Optical Thickness & unitless & 0.05 & 0.4 \\
G & Asymmetry parameter & unitless  & 0.65 & 0.99 \\
$\alpha$ &  \r{A}ngstr\"om exponent & unitless & 1 & 2 \\
SSA & Single Scattering Albedo & unitless & 0.75 & 1 \\
\hline
	   \end{tabular}}
	   \label{table_inputs2}
\end{center}
\end{table}

The MODTRAN LUTs consist on TOA radiance spectra constructed according to equation \ref{eqn:TOA} under the Lambertian assumption:
\begin{equation}
    L_{TOA} = L_0 + \frac{(E_{dir}\mu_s+E_{dif})(T_{dif}+T_{dir}) \rho }{\pi(1-S \rho)} ,
\label{eqn:TOA}
\end{equation}

where $L_0$ is the path radiance, $E_{dir/dif}$ are the direct/diffuse at-surface solar irradiance, $T_{dir/dif}$ are the surface-to-sensor direct/diffuse atmospheric transmittance, $S$ is the spherical albedo, $\mu_s$ is the cosine of \emph{SZA}, and $\rho$ is the Lambertian surface reflectance (in our case we used the conifer trees surface reflectance from ASTER spectral library~\cite{Baldridge2009711}). The atmospheric transfer functions are derived after applying the MODTRAN interrogation technique described in~\cite{guanter2009}.\\

For the emulation approach, each LUT was used to develop and evaluate the different statistical models. 
{The role of number of components has been systematically studied before~\cite{Verrelst2017}. The selection of 10 and 20 PCA components (i.e.,$\sim$100\% explained variance) was found an acceptable trade-off between accuracy and processing time. Better reconstruction of the spectral profiles can be achieved with additional components but at expenses of slower processing times.}
Further, since emulators only produce an approximation of the original model, it is important to realize that such an approximation introduces a source of uncertainty referred to as ``code uncertainty'' associated to the emulator~\cite{OHagan2006}. Therefore, validation of the generated model is an important step in order to quantify the emulator's degree of accuracy. To test the accuracy of the 500- and 2000-LUT emulators, part of the original data is kept aside as validation dataset. Various training/validation sampling design strategies are possible with the 'Emulator toolbox'.
Because of the deterministic nature of RTM data, an initial cross-validation sampling testing led to similar accuracies as one-time validation. To speed up processing time~\cite{Verrelst2016GSA}, a single single data split was therefore applied, using 70\% samples for training and the remaining 30\% for validation.

\subsection{Validation}\label{validation}

In order to show the differences between the RTM outputs and the approximation inferred by interpolation or emulation techniques, some goodness-of-fit statistics as a function of wavelength are calculated against the $n$=5000 references LUTs as generated by the RTMs. The root-mean-square-error (RMSE) and the normalized RMSE (NRMSE) [\%] (see equations \ref{eqn:rmse} and \ref{eqn:nrmse}) are calculated, both per wavelength and then averaged over all wavelengths ($\lambda$):
{
\begin{equation}
    \textit{RMSE} = \sqrt{\frac{1}{n}\sum_{i=1}^n[\f(\x_i)-\hat\f(\x_i)]^2} 
\label{eqn:rmse}
\end{equation}
\begin{equation}
    \textit{NRMSE} = \frac{100\cdot\textit{RMSE}}{\f_{max}-\f_{min}} , 
\label{eqn:nrmse}
\end{equation}
}

where $\f_{max}$ and $\f_{min}$ are respectively the maximum and minimum values of the $n$ spectra in the reference dataset. A closer inspection will be given to the most interesting results by plotting the histogram of the relative residuals ($\varepsilon_i$, in absolute terms and expressed in \%):

{
\begin{equation}
    \varepsilon_i = 100\frac{|\f(\x_i)-\hat\f(\x_i)|}{\f(\x_i)} , 
\label{eqn:residual}
\end{equation}
}

Specifically, the average relative error and the percentiles 2.5\%, 16\%, 84\% and 95.5\% will be plotted as function of wavelength.

The processing time of executing the emulator/interpolation method on the reference dataset has also been tracked. These calculations were performed in a i7-4710MP CPU at 2.5GHz with 16 GB of RAM and 64-bits operating system.

\section{Results}\label{sec:results}
In this section we will show the results of applying the emulator and interpolation methods on the described canopy and top-of-atmosphere datasets. In Section \ref{comparison}, we will show an overview of the performance of interpolation and emulation methods in terms of accuracy and computation time.  In Section \ref{closerinspection}, we will inspect in greater detail the error histograms for the best performing interpolation and emulation methods.

\subsection{Interpolation vs. emulation comparison}\label{comparison}

For both PROSAIL and MODTRAN outputs four scenarios are evaluated: training/interpolating with 500 and 2000 samples. The emulation approach is additionally tested with entering 10 or 20 components in the regression algorithm. All approaches are validated against the reference 5000 samples' LUTs.

Starting with the PROSAIL analysis, validation results and processing time is given in Table \ref{PROSAIL}. NRMSE results along the spectral range for the four scenarios are displayed in Figure \ref{PROSAIL_NRMSE}. Inspection of these four graphs suggest the following. Each of the four scenarios show approximately the same patterns, with expected higher NMRSE errors in spectral channels with lower reflectance values (e.g., bottom of Chlorophyll absorption at 680 nm and inside the water absorptions at 1440 nm and 1900 nm). The three emulation methods clearly outperform the three interpolation methods in reproducing LUT reflectance spectra. GPR is best able to reconstruct spectra with high accuracies (i.e. low NRMSE errors). KRR is second best performing, while NN performs still better than the interpolation techniques but no longer with a substantial gain in accuracy. Among the interpolation methods, linear and IDW achieve similar accuracy, particularly when using the 2000 samples LUT. Only in the near-infrared plateau (i.e., 720-1300 nm) linear interpolation obtain the lowest NRMSE errors among the interpolation techniques, similar to those obtained with NN emulation.
Thereby, results improved when more input data is involved, i.e. when the statistical models are trained with more samples, or when a denser LUT is used for interpolation. This is clearly notable when comparing the results of 2000 samples with that those of 500. A decrease in errors is especially noticeable for KRR, but also the interpolation methods lower errors with a few percents. The superiority of emulation methods can perhaps be better appreciated when considering  Table \ref{PROSAIL}: GPR trained by a 2000-LUT yielded RMSE$\boldsymbol{_\lambda}$ on the order of 10 times lower than the best interpolation method. 

PROSAIL emulation results can further improved when more components are entered in the statistical learning, as then in principle more variability is preserved. However, these improvements were not obvious in our results: when doubling the number of components from 10 to 20 hardly differences were observed for KRR and GPR. This is especially the case for the 2000 training LUT: Table \ref{PROSAIL} gives the same RMSE$\boldsymbol{_\lambda}$ results. Hence, this suggests that about 10 components are more than enough to preserve a maximal amount of information. NN appears more affected by the number of components in case of the 500 training samples: clear improvements can be observed from 1500~nm onwards. Conversely, in the visible part errors increased, which implies that the gain of adding more components is not systematic. In case of trained with 2000 samples then doubling the components did not influence at all.

When subsequently also considering processing time (see Table~\ref{PROSAIL}), then the emulation methods become even more attractive. Although the interpolation methods nearest neighbour and IDW are very fast {(below 1\% of the slowest, linear interpolation, method)}, they are not the most accurate: especially nearest neighbour is fastest but also the poorest performing. 
On the contrary, the emulation methods are not only accurate, but are also very fast. GPR processes the output spectra with a speed that is on the order of these interpolation methods. However, GPR is affected by the training size and number of components, which slows down somewhat the processing. Yet, even for the 2000 LUT and including 20 components the processing of 5000 output spectra took only a few seconds {,i.e., 2.4\% of the time spent by linear interpolation}. NN and KRR deliver spectral output still several times faster, in a fraction of a second, and this regardless of the training size. Hence, when a trade-off between accuracy and processing speed is to be made, given that NN delivers poorer accuracies, then KRR tends to become an attractive option. In all cases KRR emulated the 5000 spectra {a factor 100 faster than linear interpolation (0.3 seconds)}, and this with second-best accuracies.

\begin{figure*}[!ht]
 \centering \small
 \IG[width=0.495\linewidth]{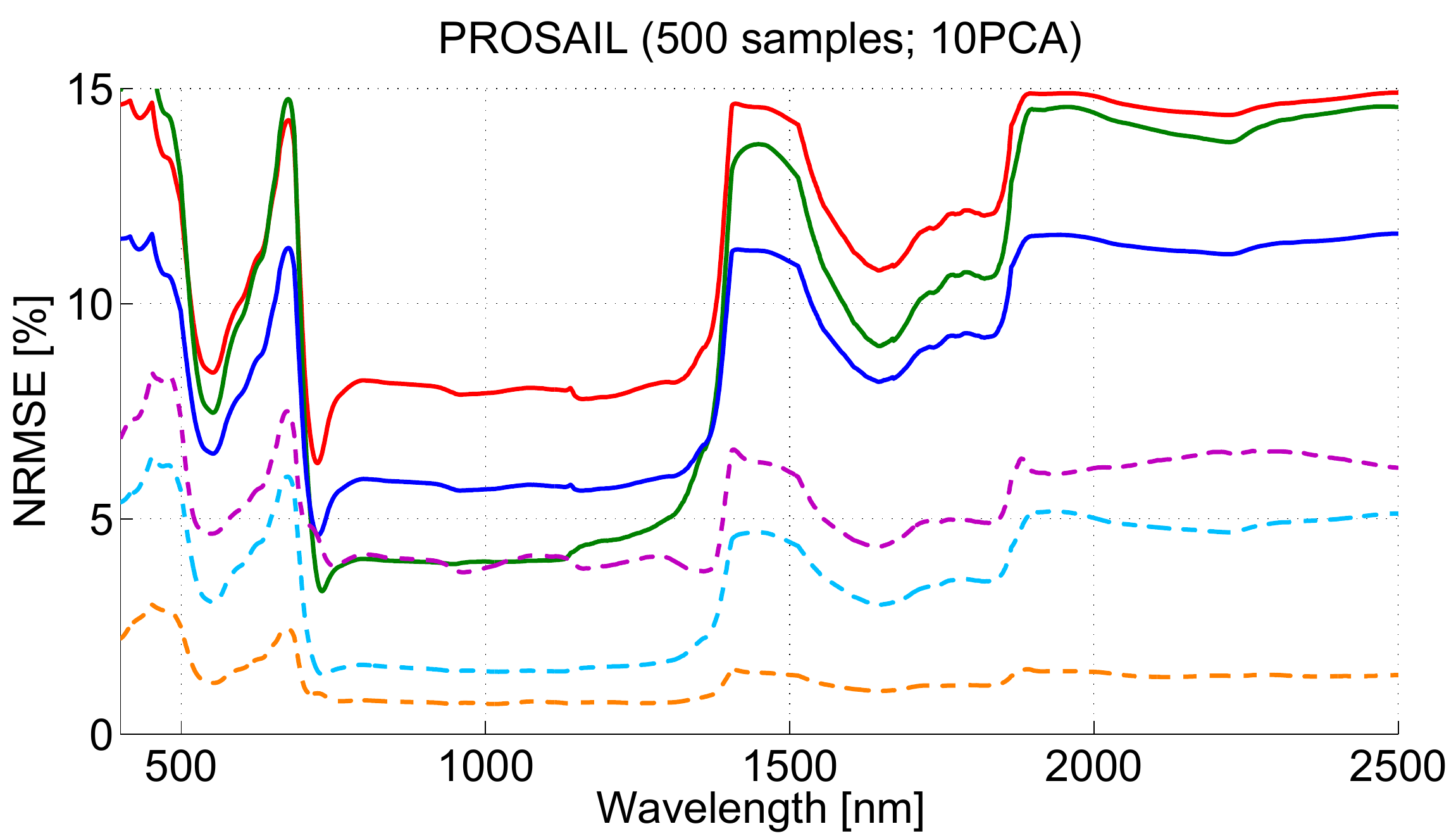}
  \IG[width=0.495\linewidth]{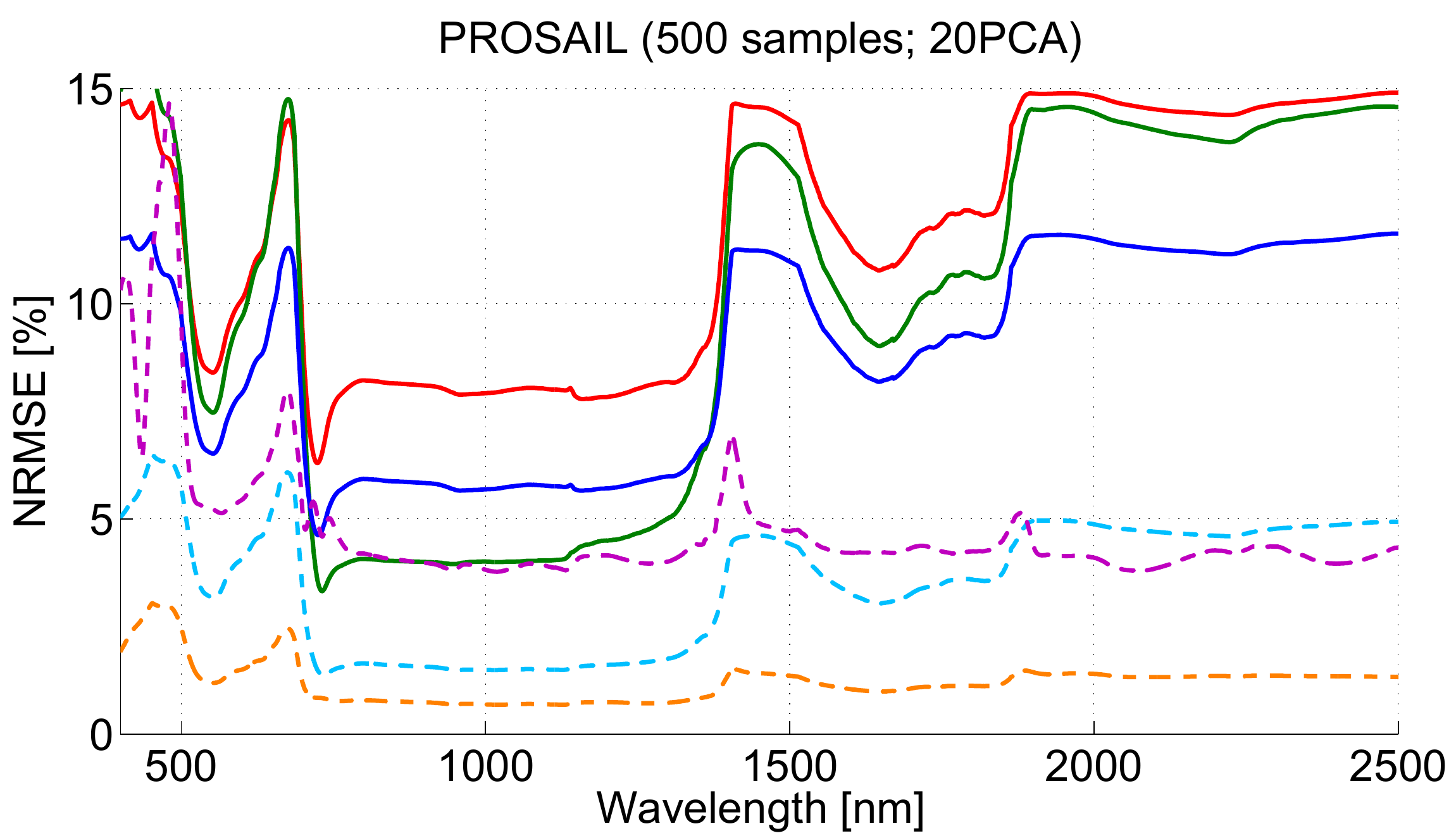} \\
 
  \IG[width=0.495\linewidth]{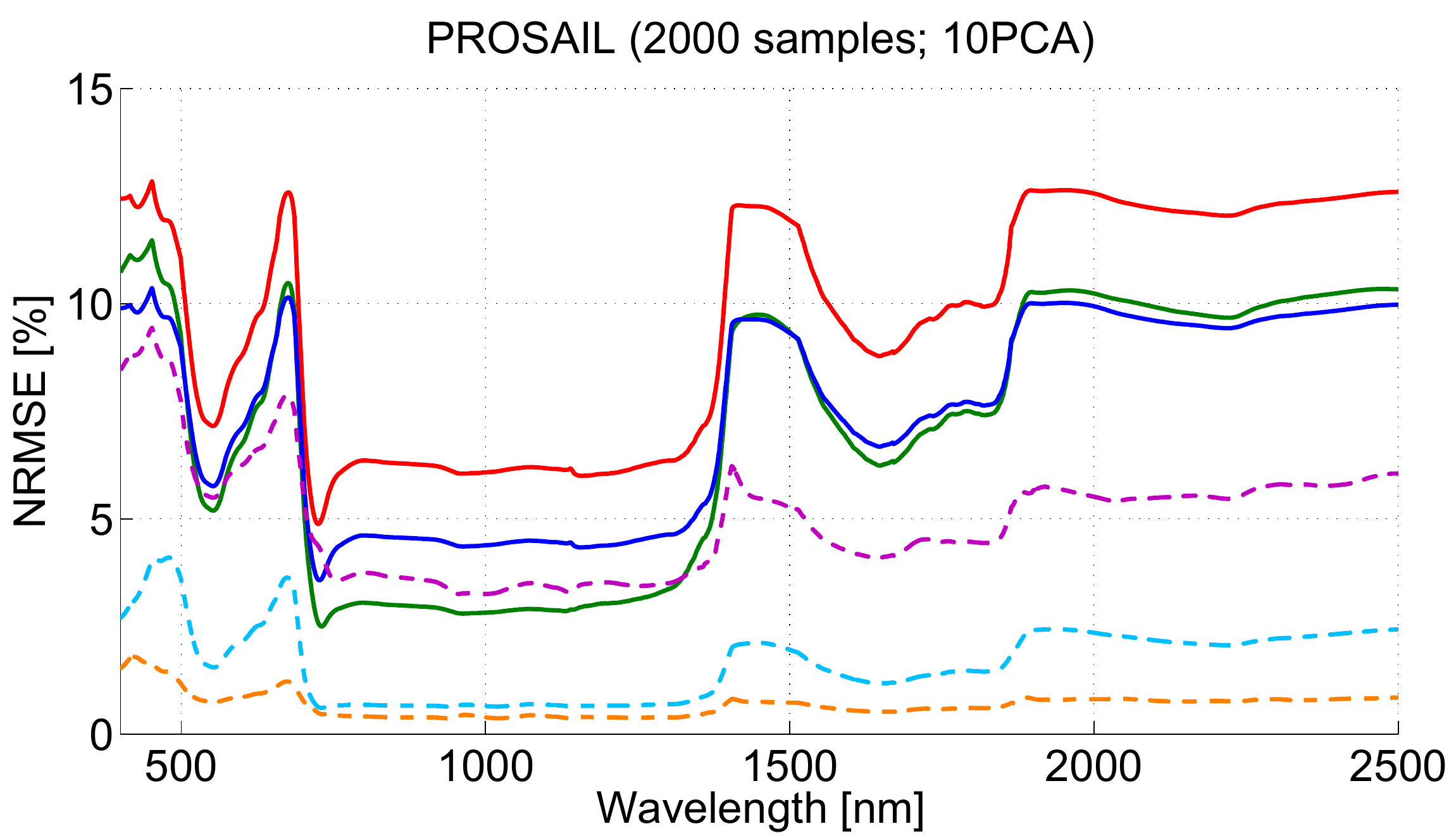}
  \IG[width=0.495\linewidth]{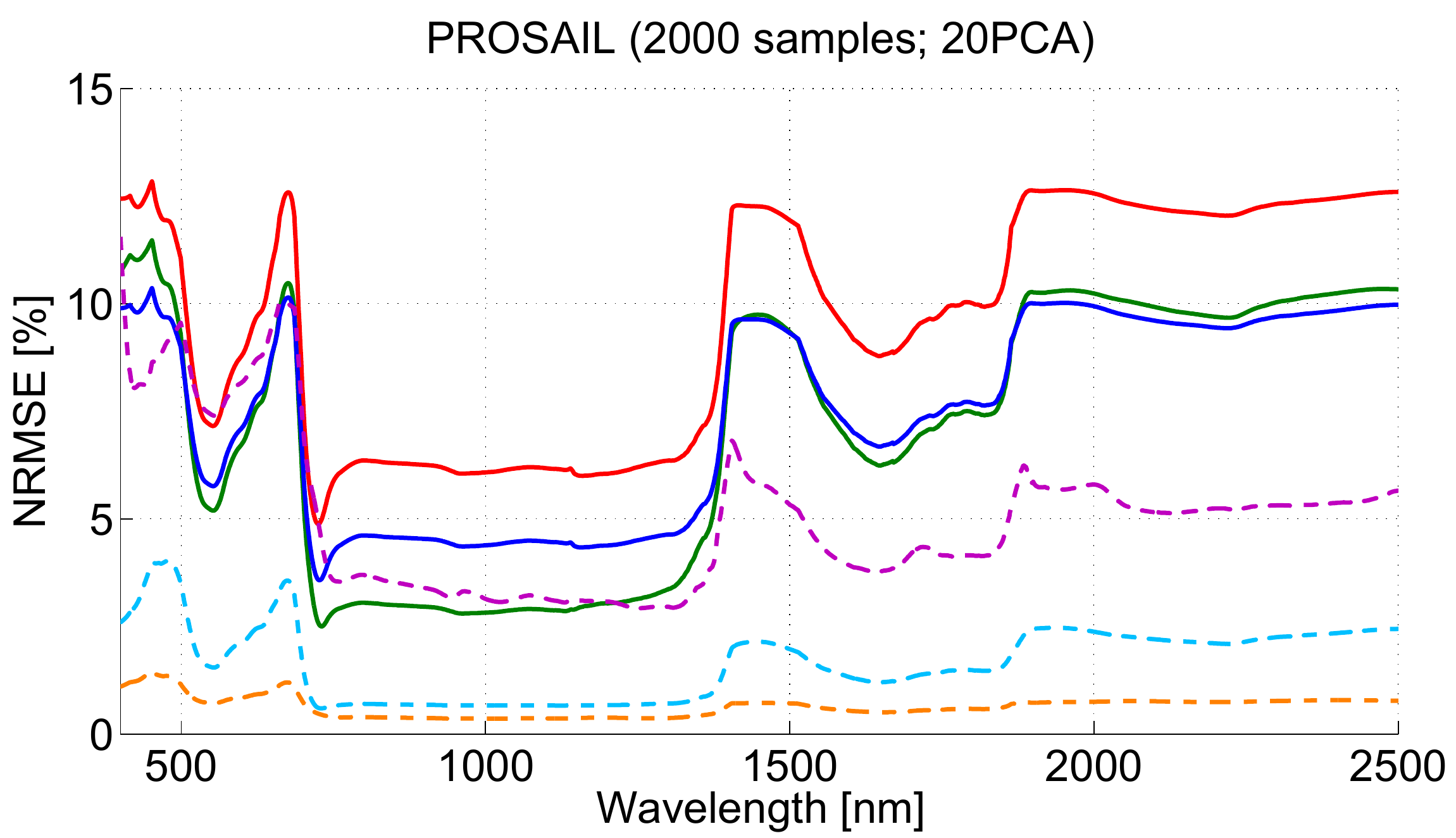} \\
  
  \IG[width=0.48\linewidth]{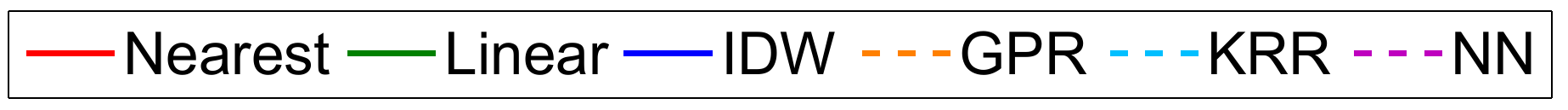}

     \caption{PROSAIL interpolation vs emulation results. {Note that the number of PCA components refers only to the emulator methods since no dimensionality reduction is applied in interpolation.}}
 \label{PROSAIL_NRMSE}
\end{figure*}

\begin{table}[!ht]
\centering \small
 \caption{PROSAIL interpolation and emulators validation results against 5000 LUT reference dataset  (RMSE$\boldsymbol{_\lambda}$, NRMSE$\boldsymbol{_\lambda}$) and processing time (s: seconds).}
     \resizebox{\linewidth}{!}{ 
 \begin{tabular}{lcccccc} \toprule 
 {\bf Method} & \multicolumn{2}{c}{\bf RMSE$\boldsymbol{_\lambda}$} & \multicolumn{2}{c}{\bf NRMSE$\boldsymbol{_\lambda}$ (\%)} & \multicolumn{2}{c}{\bf CPU (s) } \\ 
 {\bf LUT training size:} &  {\bf 500} &  {\bf 2000} &  {\bf 500} &  {\bf 2000} &  {\bf 500} &  {\bf 2000} \\

\midrule
{\bf Interpolation:} &&&&&& \\
 - Nearest & 0.051 & 0.042 &  11.60 & 9.61 & 0.2 & 0.7 \\
 - Linear & 0.042 & 0.030 & 9.99 & 7.05  & 62 & 171 \\
 - IDW & 0.039 & 0.032  & 8.86 & 7.44  & 0.5 & 1.2 \\
{\bf Emulation 10PCA}: &&& &&&\\
 - GPR & 0.005 & 0.003 & 1.23 & 0.68 &  0.7 & 2.2 \\
 - KRR & 0.015 & 0.007 & 3.56 & 1.67 &   0.1 & 0.2 \\
 - NN & 0.024 & 0.022 & 5.33 & 5.00 &  0.2 & 0.2 \\
{\bf Emulation 20PCA}: &&& &&&\\
 - GPR & 0.005 & 0.003 & 1.21 & 0.64 &  0.9 & 4.2 \\
 - KRR & 0.015 & 0.007 & 3.54 & 1.67 &  0.3 & 0.3 \\
 - NN &  0.021 & 0.022 & 4.74 & 5.04 &  0.2 &  0.2\\
\bottomrule
  \end{tabular}}
  \label{PROSAIL}
\end{table}

The same analysis has been repeated with the MODTRAN LUTs (Table \ref{MODTRAN} and Figure \ref{MODTRAN_NRMSE}). NRMSE results are now plotted in logarithmic scale in order to better visualize differences between the various interpolation/emulation methods and the wide range of error values between inside and outside atmospheric absorption bands. Similar patterns as the PROSAIL results are obtained, yet some differences should be remarked.
GPR and secondly KRR are again clearly top performing LUT parameter sampling methods. Table \ref{MODTRAN} indicates that KRR and especially GPR yielded RMSE$\boldsymbol{_\lambda}$ results more than 10 times lower than best performing interpolation method. However, NN performs now on the order of the interpolation methods. Regarding the interpolation methods, linear interpolation now systematically outperformed the other two methods but still errors are nearly one order of magnitude higher than the GPR and KRR emulators.
Moving from 500 to 2000 training samples did not lead to significant improvements. Table \ref{MODTRAN} suggests that only for KRR a substantial gain in accuracy was achieved. The same holds for adding more components into the emulators: although some small improvements can be obtained with more components, e.g. as is noticeable for GPR, overall the gain in accuracy is modest.

Also regarding processing time similar trends emerged as for PROSAIL: all three emulation methods produced the spectral output very fast, with NN and KRR delivering the 5000 MODTRAN-like spectra in a fraction of a second {(a factor $<$1\% when compared against linear interpolation)}. GPR suffered somewhat from adding more samples and components, leading to a slightly slower emulation method in case of 2000 LUT and trained with 20 components: the output spectra is again produced in a few seconds.

\begin{figure*}[!ht]
 \centering \small
 \IG[width=0.495\linewidth]{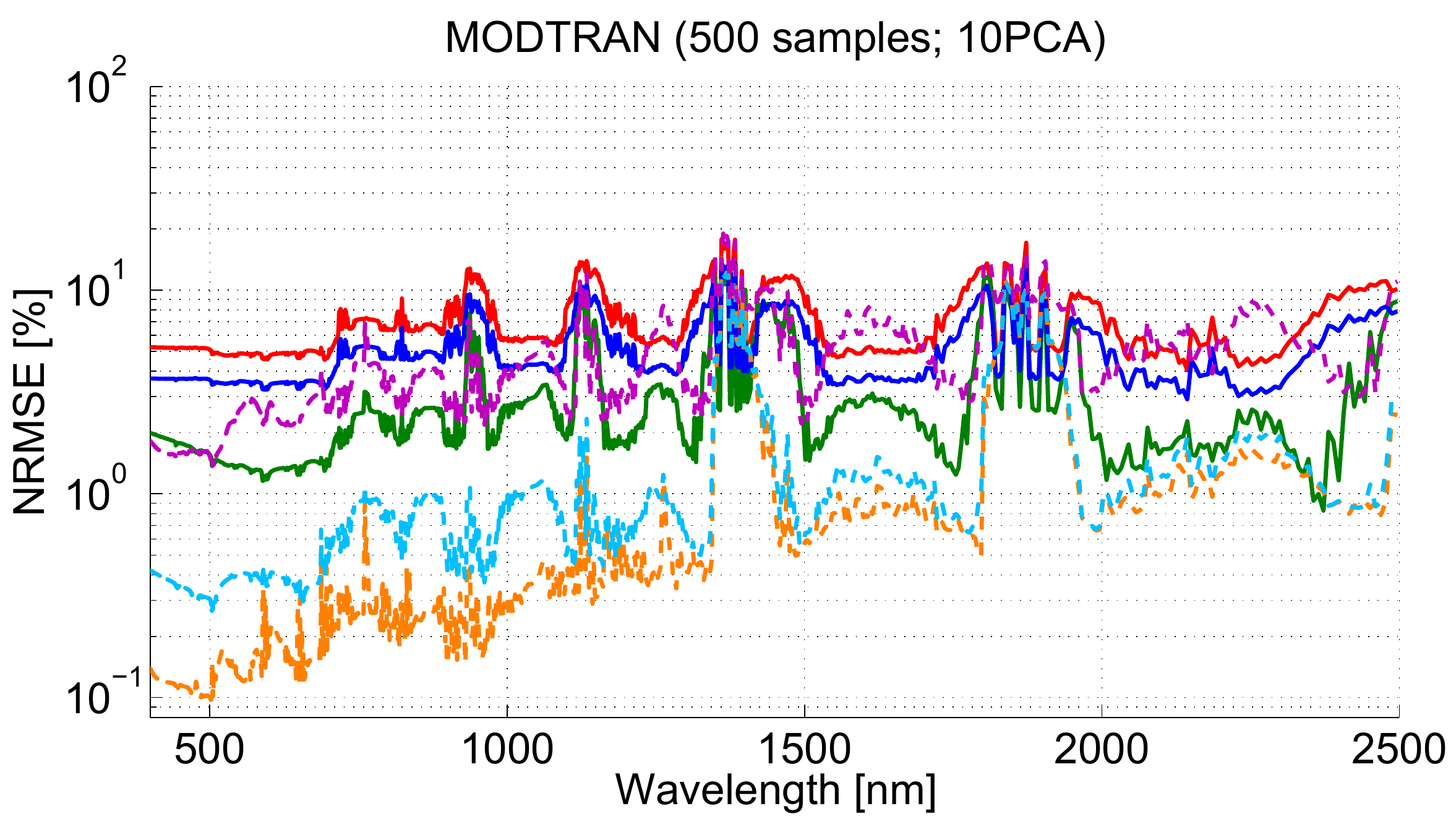}
  \IG[width=0.495\linewidth]{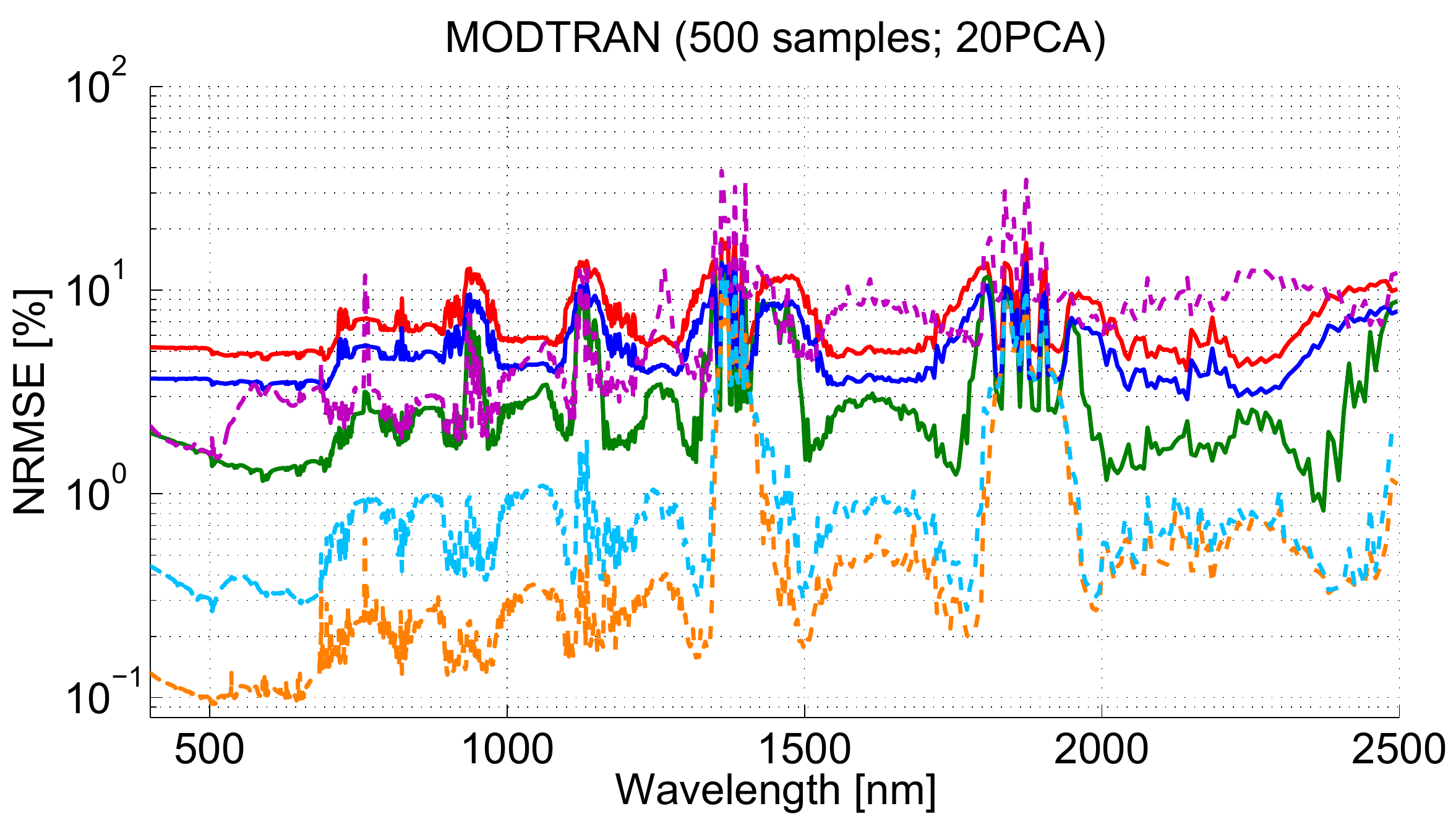} \\
 
  \IG[width=0.495\linewidth]{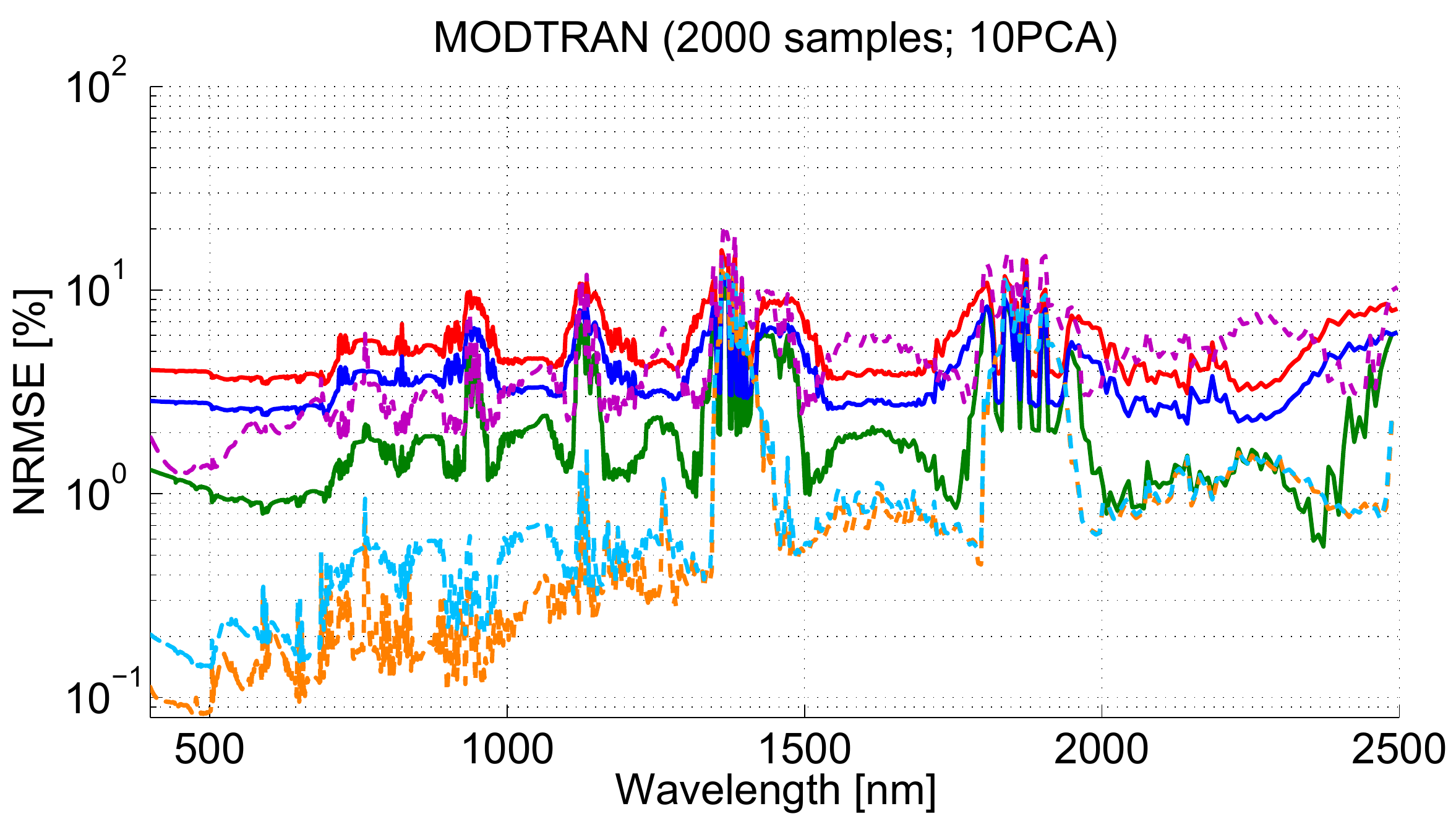} 
  \IG[width=0.495\linewidth]{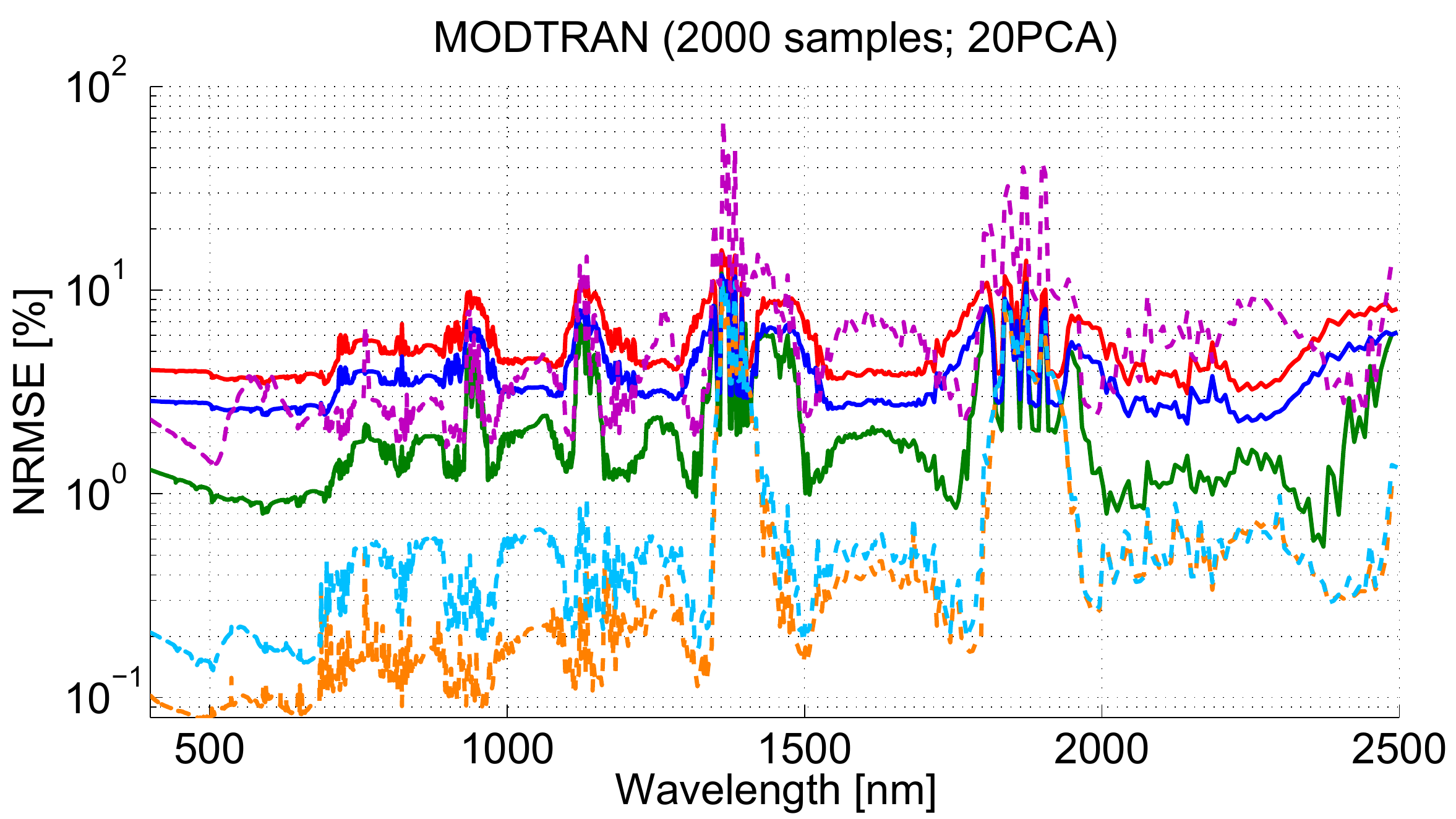} \\
\IG[width=0.48\linewidth]{figures/vicen6}

     \caption{MODTRAN interpolation vs emulation results. {Note that the number of PCA components refers only to the emulator methods since no dimensionality reduction is applied in interpolation.}}
 \label{MODTRAN_NRMSE}
\end{figure*}

\begin{table}[!ht] 
\centering \small
 \caption{MODTRAN interpolation and emulators validation results against 5000 LUT reference dataset  (RMSE$\boldsymbol{_\lambda}$, NRMSE$\boldsymbol{_\lambda}$) and processing time (s: seconds).}
     \resizebox{\linewidth}{!}{ 
 \begin{tabular}{lcccccc} \toprule 
 {\bf Method} & \multicolumn{2}{c}{\bf RMSE$\boldsymbol{_\lambda}$} & \multicolumn{2}{c}{\bf NRMSE$\boldsymbol{_\lambda}$ (\%)} & \multicolumn{2}{c}{\bf CPU (s) } \\ 
 {\bf LUT training size:} &  {\bf 500} &  {\bf 2000} &  {\bf 500} &  {\bf 2000} &  {\bf 500} &  {\bf 2000} \\

\midrule
{\bf Interpolation:} &&&&&& \\
 - Nearest & 1.160 & 0.896 &  7.58 & 5.83 & 0.3 & 0.8 \\
 - Linear & 0.386 & 0.265 & 2.84 & 1.98  & 68 & 183 \\
 - IDW & 0.837 & 0.630  & 5.47 & 4.12 & 0.5 & 1.3 \\ 
{\bf Emulation 10PCA}: &&& &&&\\
 - GPR & 0.037 & 0.031 & 0.65 & 0.59 & 0.5 & 2.0 \\
 - KRR & 0.100 & 0.054 & 1.01 & 0.73 & 0.1 & 0.2  \\
 - NN & 0.488 & 0.406 & 4.13 & 3.62 & 0.1  & 0.1 \\
{\bf Emulation 20PCA}: &&& &&&\\
 - GPR & 0.029 & 0.022 & 0.43 & 0.37 & 1.2 & 4.7 \\
 - KRR & 0.094 & 0.051 & 0.84 & 0.56 & 0.1 & 0.2 \\
 - NN &  0.487 & 0.466 & 4.83 & 4.40 & 0.1 & 0.1 \\
\bottomrule
  \end{tabular}}
  \label{MODTRAN}
\end{table}

\subsection{Closer inspection of best performing results}\label{closerinspection}

Having observed the general trends of the interpolation and emulation methods, in this section we will inspect a few methods in more detail. Specifically, the histograms of the relative residuals (in absolute terms) for the best performing interpolation and emulation methods, i.e. linear and GPR, are plotted in Figures \ref{residualsPROSAIL} and \ref{residualsMODTRAN}. Thereby, the linear interpolation method is shown as obtained with a 2000-LUT, whereas GPR is shown as trained with only a 500-LUT and 10 PCA components. Interestingly, although the emulator method is not presented in its optimized configuration, already a substantial gain in accuracy as compared to the optimized linear interpolation method is achieved.  

To fully appreciate the predictive power of the emulator methods, we start by analyzing the results on the PROSAIL residuals in Figure \ref{residualsPROSAIL}. 
We can observe how the GPR emulator obtains relative errors that, on average, are a factor 5-10 lower than those obtained with the linear interpolation (see mean values on the solid line). These error differences between the GPR emulator and linear interpolation methods are still maintained on both the lower and higher part of the histogram (see 2.5\% and 84\% percentiles, where the reconstruction errors are lower/higher respectively).

\begin{figure}[!ht]
	\centering \small
	\IG[width=\linewidth]{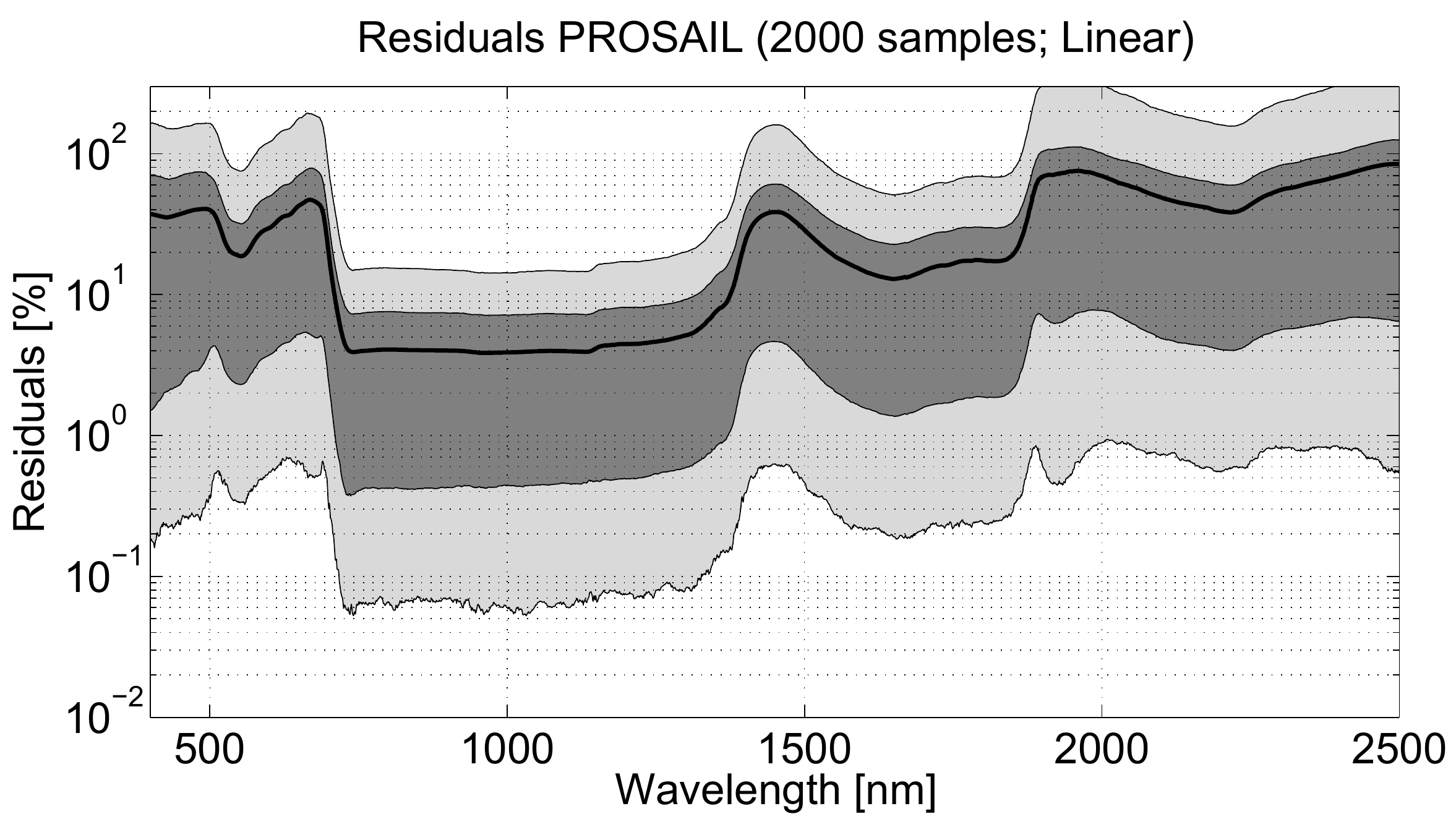} \\
	\IG[width=\linewidth]{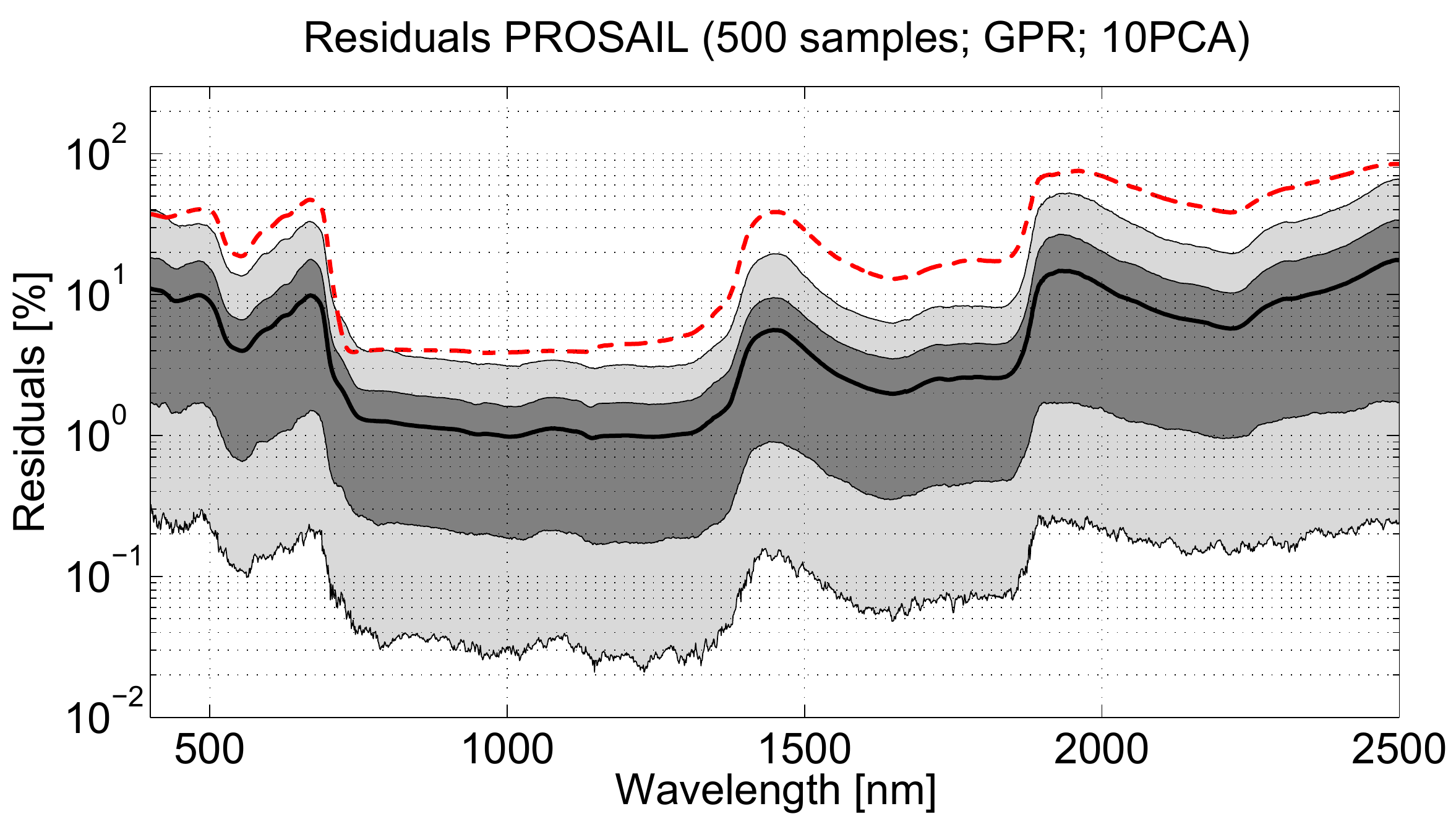}\\
	\IG[width=0.8\linewidth]{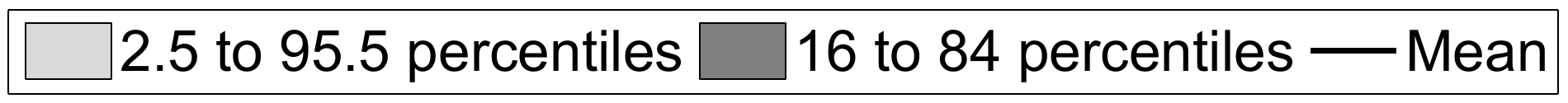}
	\caption{Histogram statistics of PROSAIL relative residuals (in absolute terms)(\%) for 2000-LUT interpolation linear ({top}) and 500-LUT GPR ({bottom}). To ease the comparison, the mean residual for linear interpolation is added on top of the GPR residuals (red dashed line).}
	\label{residualsPROSAIL}
\end{figure}

We continue by analyzing the results on the MODTRAN residuals in Figure \ref{residualsMODTRAN}.
As previously observed in NRMSE values (see Figure \ref{MODTRAN_NRMSE} and Table \ref{MODTRAN}), the reconstruction errors with GPR emulators obtains the best performance. The residual errors are in this case a factor 2-10 lower than using linear interpolation for both the lowest and highest errors in the histogram and in most part of the spectrum ($<$1800~nm).

\begin{figure}[!ht]
	\centering \small
	\IG[width=\linewidth]{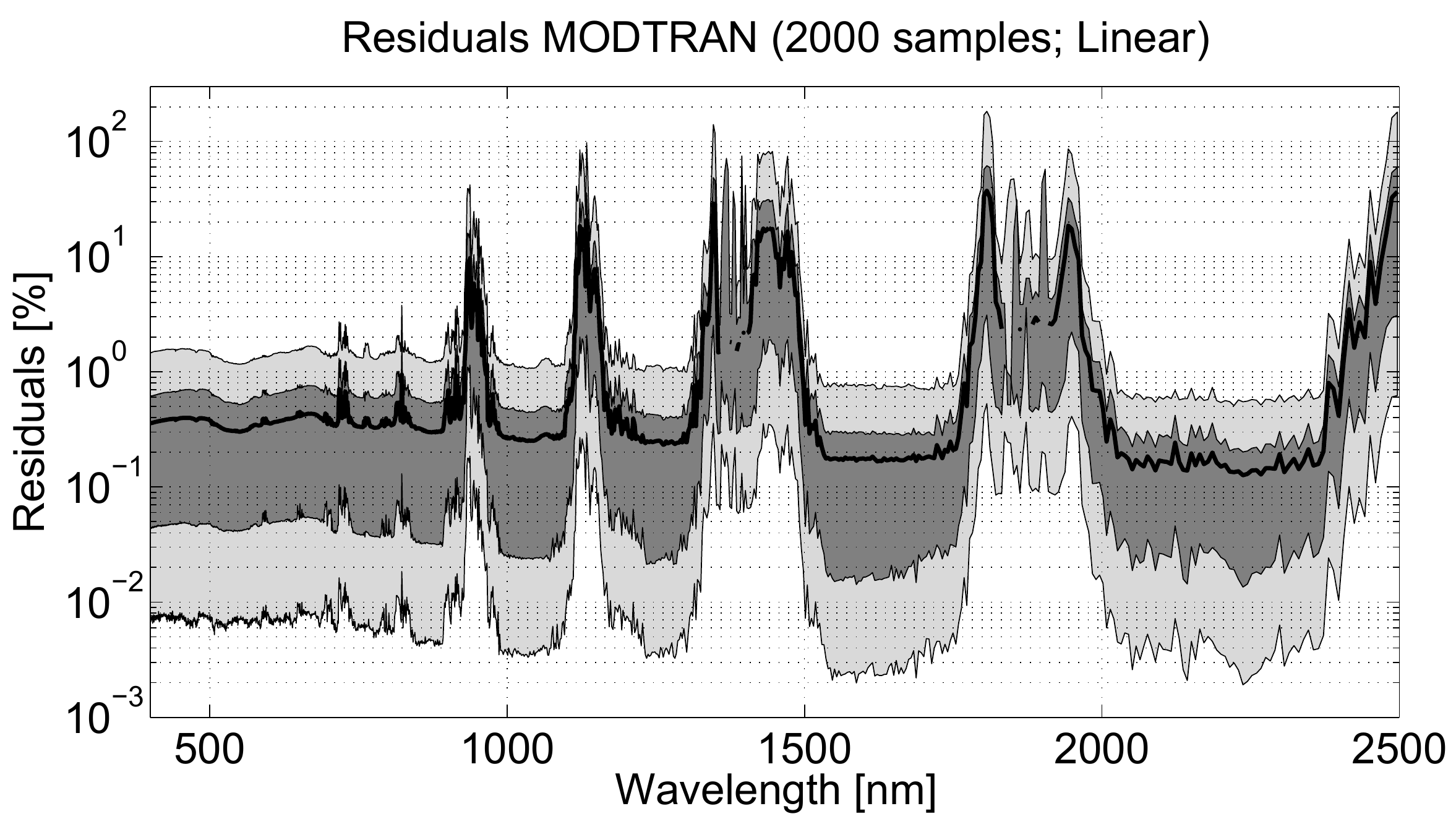}
	\IG[width=\linewidth]{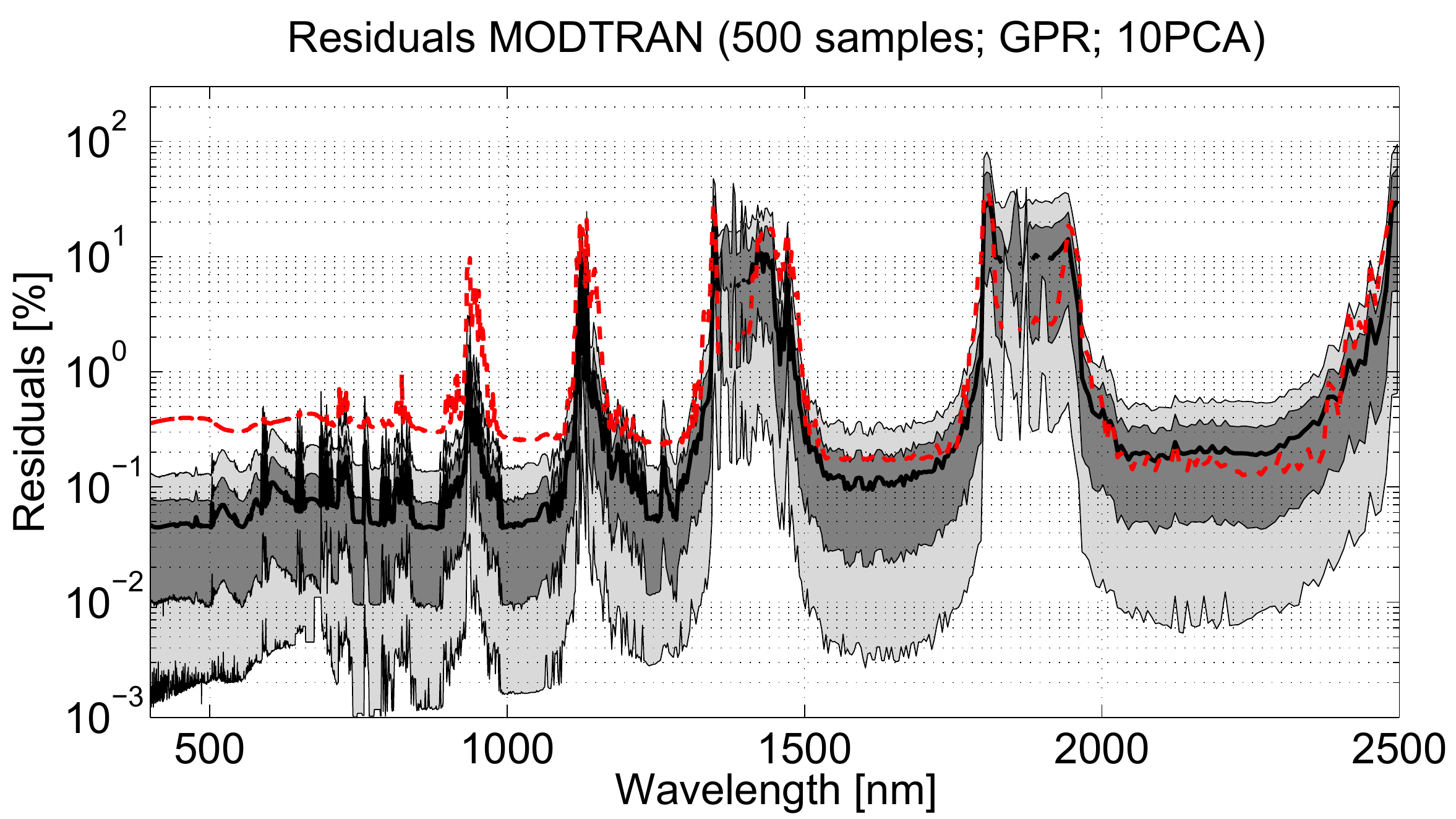}\\
	\IG[width=0.8\linewidth]{figures/vicen13.pdf}
	\caption{Histogram statistics of MODTRAN relative residuals (in absolute terms)(\%) for 2000-LUT interpolation linear ({top}) and 500-LUT GPR ({bottom}). To ease the comparison, the mean residual for linear interpolation is added on top of the GPR residuals (red dashed line).}
	\label{residualsMODTRAN}
\end{figure}

\section{Discussion}\label{sec:discussion}
Advanced RTMs are widely used {in various remote sensing applications} such as (i) atmospheric correction (e.g.,~\cite{Richter1996}), (ii) inversion of vegetation properties (see \cite{Verrelst2015b} for a review), (iii) sensitivity analysis \cite{Verrelst2015a}, and (iv) scene generation \cite{Verhoef2012}. Because advanced RTMs take long processing time, LUT interpolation is commonly used to sample the input parameter space and to infer an approximation of the RTM outputs~\cite{guanter2009}.
Although various interpolation techniques are standard practice in remote sensing applications and computer vision, in this work we challenged these techniques by comparing them against statistical learning methods, i.e., \emph{emulation}. 
The basic principle of emulation is using a sparse LUT to train machine learning methods so that the trained statistical model is able to reproduce the spectral output given unseen parameter combinations. According to this principle, the emulation technique can be considered to function similarly as interpolation methods, but based on statistical learning. 

To ascertain the predictive power of both interpolation and emulation, two experiments were conducted: one for surface reflectance as generated by PROSAIL, and another for TOA radiance data as generated by MODTRAN. For both experiments the interpolation and emulation methods were tested against a common reference dataset of 5000 simulations. Despite TOA radiance data being much more irregular and spiky (due to narrow atmospheric absorption regions) than surface reflectance, common trends were observed in the majority  of cases. They are summarized below:

\begin{itemize}
\item  In all experiments, the three tested emulation methods produced substantially more accurately spectral outputs than the three tested interpolation techniques. Particularly, GPR was by far the most accurate emulation method with errors on the order of 5-10 times lower than the linear interpolation method and at a fraction of the time with respect linear interpolation. KRR was the second most accurate emulation method, with reconstruction errors 2-5 times lower than the best performing interpolation method. GPR is thus clearly the preferred method given accuracy and fast processing. At the same time, KRR run still much faster than GPR, i.e., producing 5000 spectra in 0.3 seconds, which is faster than any interpolation method behind while reaching superior accuracies.  
\item  Regarding the emulation methods, {GPR and KRR emulate output spectra not only perform more accurately but also tend to be more stable than NN (e.g. Figure \ref{PROSAIL_NRMSE},  in the visible part). Having more parameters than KRR/GPR to adjust, training the NN emulator can be more complicated, and thus requires a larger number of training samples to avoid overfitting and to obtain a good prediction. In addition, an increasing number of components can make the problem harder in terms of adjusting the NNs parameters. Conversely, for GPR and KRR clearly} some little extra gain can be achieved by training with more samples or by adding more PCA components in the regression model. However, this slows down processing time, especially in case of GPR (see also~\cite{Verrelst2017} for a discussion on GPR emulation).
\item  When comparing the PROSAIL emulation results with those of MODTRAN, e.g., as in the histogram statistics (Figures \ref{residualsPROSAIL} and \ref{residualsMODTRAN}), it is noteworthy that the spectrally smoother TOC reflectance is reconstructed with lower accuracy than the more irregular TOA radiance. Although the parameter space of both LUTs consists of 6 variables, the discrepancy can be explained that the 6 PROSAIL variables exert more influence, in relative terms, on the reflectance data than the 6 MODTRAN variables on the TOA radiance data, implying more variability to construct the reflectance spectra. This has been observed before with a global sensitivity analysis~\cite{Verrelst2016GSA}.   
\item  Regarding the interpolation methods, while performing substantially poorer than emulation, linear interpolation is the most accurate method. This is particularly the case when increasing the LUT size. However, it is also the most computationally intensive interpolation method, i.e., on the order of few minutes for 5000 spectra. This is mostly due to the exponential increase of memory usage in construction of the implicit Delaunay triangulation~\cite{Barber1996469}. 
\end{itemize}

The superiority of KRR and especially GPR emulation is most noteworthy. The strength of GPR statistical learning was already demonstrated in previous studies with application of machine learning algorithms in various fields \cite{OHagan94,Oakley2004,Conti2009}, yet in fact any machine learning regression algorithm can function as emulator.
Moreover, since the accuracy and run-time of these methods depend on the size of training LUT and number of components, these methods can be further optimized in view of balancing between accuracy and processing speed. For instance, it is likely that GPR can still deliver excellent accuracies with a smaller training LUT or less components, which would imply a faster processing. As addressed before \cite{Verrelst2017}, it requires some iterations to deduce an optimized accuracy-speed trade-off.  

Hereby, another point to be remarked is that emulation requires the additional effort of a training step and validation of the model. This implies that a sparse LUT is always required to train an emulator, i.e. to finding the best hyperparameters that optimize its performance. Two issues have an impact on finding these optimal parameters: (1) the method for tuning the hyperparameters and (2) the size of the training dataset.
Regarding the first issue, and in the specific case of the implemented GPR, {the maximum log-likelihood method was adopted to find the best set of hyperparameters. Nevertheless, other alternative procedures are often employed with similar success and adoption. Examples include random sampling~\cite{Bergstra12}, the Nelder-Mead method (aka downhill simplex)~\cite{Nelder65}, Bayesian optimization~\cite{Nelder65,Gutmann15,Mockus89,Snoek12}, and many flavours of derivative-free optimization approaches such as stochastic local search, simulated annealing or evolutionary computation~\cite{Kirkpatrick83,OMCMC}. Other approaches consider Monte Carlo methods~\cite{Robert04,MartinoMH17,MartinoJesse1}, which search in a portion of the space according to the posterior distribution of the hyper-parameters given the observed data.}
With respect to the training dataset, we found in our examples that typically about 500 samples according Latin hypercube sampling should suffice. The training and model validation step requires some additional processing time as compared to interpolation. In case of KRR this is in the order of seconds, but NN and GPR that can take longer depending on the complexity of training setup (number of samples and components). Nevertheless, the training phase is to be done only one time; the generated emulator model can afterwards replace the LUT interpolation in the processing chain. This approach of replacing a LUT interpolation by an emulator may lead not only to a more accurate processing, but likely also faster and less computationally demanding.

With respect to interpolation methods, in this paper we focused on exploring the accuracy of the most widely used methods working on scattered LUT data. From the considered methods, linear interpolation achieve the higher accuracy. However, this method is typically limited to low dimensions ($<$6) of the input space due to its exponential demands of memory usage and computation time.
Other more advanced interpolation methods (not studied here) exist for gridded LUT data. Among them, piece-wise cubic splines interpolation~\cite{Bartels} can achieve high accuracies but at expenses of an increase of computation time with respect linear interpolation. Also, cubic splines requires gridded LUTs with at least four points in each dimension (e.g., 4096 LUT samples for 6 dimensions), which implies both an increase of memory usage and computation time to generate and interpolate the resulting LUTs. 
Sibson method is another advanced and accurate interpolation algorithm~\cite{sibson} that can have very fast implementations for low dimension spaces (see e.g. \cite{Park2006}). However, the extension of Sibson interpolation in high dimensional input variable spaces ($>$6) is likely not to be effective due to its exponential growth in memory consumption. 
In turn, the emulation approach only requires a small LUT for training, which in principle can be developed for any number of variables. It remains yet to be studied, how adding more variables to the LUT parameter space affects the accuracy of the emulator. This will also depend on the role these variables play in driving the spectral output \cite{Verrelst2017}.

Altogether, considering all strengths and weaknesses of both interpolation and emulation, this study leads us concluding that the emulation technique can become an attractive alternative to interpolation in sampling a LUT parameter space. 
Based on the results presented here, it is foreseen that relying on emulation rather than interpolation will lead to more accurate, and typically faster querying through a LUT parameter space.
This bears consequences in various RTM-based processing applications in which high accuracy and fast run-time is needed, e.g., in scene generation~{\cite{Ientilucci2003110,Tenjo2017,Verrelst2017}}, in atmospheric correction procedures, in LUT-based inversion of biophysical parameters \cite{Rivera13a,Verrelst14}, {in hyperspectral target detection~\cite{Matteoli20111343}} or {in instrument performance modelling~\cite{Kerekes2005571}.}
Further studies are required to consolidate whether emulation techniques are always to be trusted as functioning more accurately than interpolation techniques.

\section{Conclusions}\label{sec:conclude}
Computationally expensive RTMs are commonly used in various remote sensing applications. Because these RTMs take long processing time, the common practice is develop one time a LUT and then making use of interpolation techniques to sample the LUT parameter space. However, the question arose whether these interpolation techniques are most accurate. This work proposed to use emulation, i.e., statistical learning, as an alternative to interpolation. Two experiments were conducted to ascertain the accuracy in delivering spectral outputs of both techniques: one for TOC reflectance as generated by PROSAIL, and another for TOA radiance data as generated by MODTRAN. The interpolation and emulation methods were evaluated against a reference LUT of 5000 simulations, leading to the following results:
(1) in all experiments the emulation methods clearly produced output spectra more accurately than the tested interpolation techniques.
(2) GPR reproduced RTM output spectra up to ten times more accurately than interpolation methods and this with a speed that is {$<$5\% of the linear interpolation method}, i.e. in mere seconds. KRR was the second most accurate method, and this emulator is extremely fast: 5000 spectra were produced in a fraction of a second. Hence, KRR shows an attractive trade-off between accuracy and computational time.
(3) Regarding emulation, some little more gain can be achieved by training with more samples or adding more PCA components in the regression model. However, for GPR this is at the expense of somewhat slowing down processing. It is thus concluded that emulation methods offer a better alternative in computational cost and accuracy than traditional methods based on interpolation to sample a LUT parameter space.

Future work will aim to include GPR emulators as an alternative to the current LUT interpolation methods implemented in the FLEX end-to-end mission simulator \cite{Vicent2016,Tenjo2017}. This will likely reduce the computation time for the generation of synthetic scenes, which will extend the current FLEX simulator capabilities to perform sensitivity analysis for various leaf/canopy and atmospheric conditions. In addition, two research lines are currently being explored to improve the accuracy of interpolation/emulation methods. On the one hand, we are optimizing the LUT nodes distribution in order to reduce LUT size while increasing the accuracy of interpolation methods: On the other hand, we can improve the accuracy of statistical methods to generate multi-output emulators by using advanced machine learning methods, deep learning algorithms or combining emulators e.g. for different wavelength regions. Both research lines will turn into more efficient sampling methods, reducing both computation burden and RTM-output reconstruction error.

\section{Bibliography}\label{sec:bib}
\begin{IEEEbiography}[{\IG[width=1in,height=1.25in,keepaspectratio]{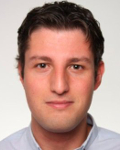}}]
{Jorge Vicent Servera} received the B.Sc. degree in physics from the University of Valencia, Spain, in 2008; the M.Sc. in physics from the EPFL, Switzerland, in 2010; and the PhD. in remote sensing from University of Valencia, Spain, in 2016. Since November 2017, he has been with Magellium in the Earth Observation department, Toulouse, France as a R\&D Engineer. He is currently involved in developing the Level-2 processing chain for ESA's FLEX mission. His research interests include the modelling of Earth Observation satellites, system engineering, radiative transfer modelling, atmospheric correction and hyperspectral data analysis.
\end{IEEEbiography}

\begin{IEEEbiography}[{\IG[width=1in,height=1.25in,keepaspectratio]{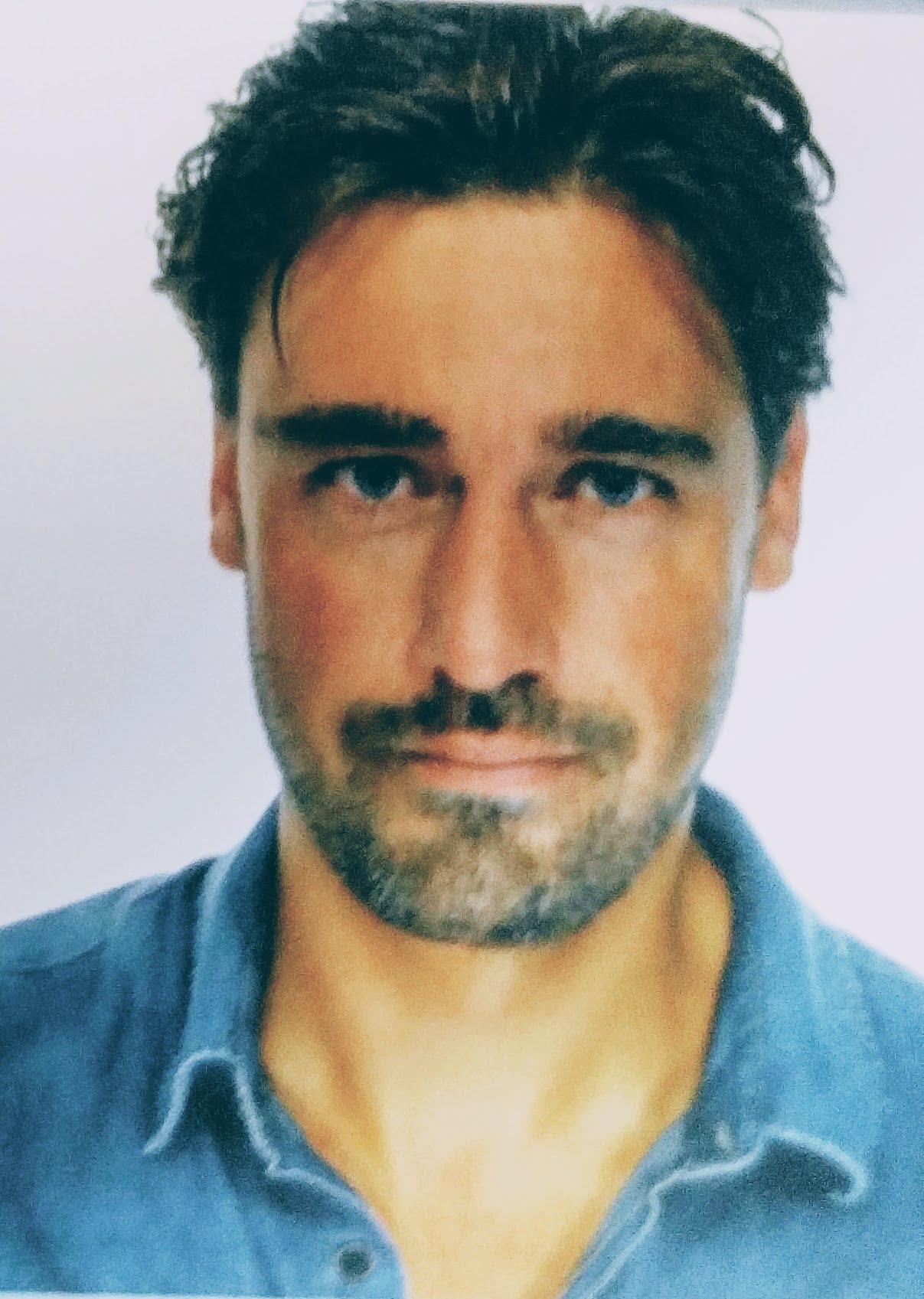}}]
{Jochem Verrelst} received the M.Sc. degree in tropical land use and in geo-information science both in 2005 and the Ph.D. in remote sensing in 2010 from Wageningen University, Wageningen, Netherlands. His dissertation focused on the space-borne spectrodirectional estimation of forest properties. Since 2010, he has been involved in preparatory activities of FLEX. His research interests include retrieval of vegetation properties using airborne and satellite data, canopy radiative transfer modeling and emulation, and hyperspectral data analysis. He is the founder of the ARTMO software package. In 2017 he received a H2020 ERC Starting Grant (\#	755617) to work on the development of vegetation products based on synergy of FLEX and Sentinel-3 data. 
\end{IEEEbiography}

\begin{IEEEbiography}[{\IG[width=1in,height=1.25in,keepaspectratio]{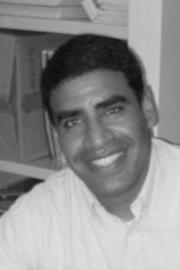}}]
{Juan Pablo Rivera-Caicedo} received his B.Sc. degree in agricultural engineering from the University National of Colombia and University of Valle, Cali, Colombia, and in 2001 his Master degree in irrigation engineering from the CEDEX-Centro de Estudios y Experimentaci\'on de Obra P\'ublicas, Madrid, Spain, in 2003. In 2011 he received a MSc and in 2014 his Ph.D. degree in Remote sensing at the University of Valencia, Valencia, Spain. Since January 2011, he has been a member of the Laboratory for Earth Observation, Image Processing Laboratory, University of Valencia, Spain. And since 2016 he works at Concejo Nacional de Ciencia y Tecnologia - CONACYT in M\'exico in the program: Catedras-Conacyt. Currently, he is involved in preparatory activities of the Fluorescence Explorer. His research interests include retrieval of vegetation properties using airborne and satellite data, leaf and canopy radiative transfer modeling and hyperspectral data analysis.
\end{IEEEbiography}

\begin{IEEEbiography}[{\IG[width=1in,height=1.25in,keepaspectratio]{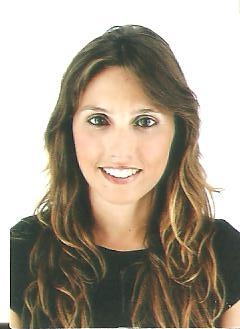}}]
{Neus Sabater} received the B.Sc. degree in physics in 2010 and M.Sc. degree in remote sensing in 2013 both from Universitat de Val\`encia, Valencia, Spain. She also received the award of the University of Valencia to the best student records in the M.Sc. of remote sensing (2012-2013). Since August 2012, she has been involved in the activities of the Laboratory for Earth Observation at the Image Processing Laboratory, University of Valencia, as a research technician. Main activities during this period were related to the development of the preparatory activities of the FLEX mission. From 2013, she was awarded a PhD scholarship from the Spanish Ministry of Economy and Competitiveness, associated to the Ingenio/Seosat Spanish space mission. Main research and personal interests include atmospheric correction, atmospheric radiative transfer, meteorology and hyperspectral RS.
\end{IEEEbiography}

\begin{IEEEbiography}
[{\IG[width=1in,height=1.25in,clip,keepaspectratio]{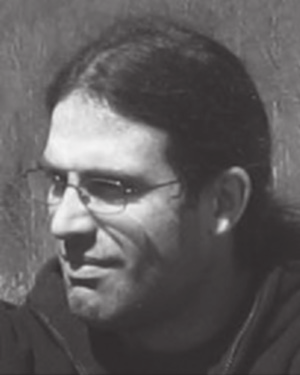}}]{Jordi Mu{\~n}oz-Mar\'i} (M'11) received a B.Sc. degree in Physics (1993), a B.Sc. degree in Electronics Engineering (1996), and a Ph.D. degree in Electronics Engineering (2003) from the Universitat de Val\`encia. He is currently an Associate Professor in the Electronics Engineering Department at the Universitat de Val\`encia, where teaches Electronic Circuits and, Programmable Logical Devices, Digital Electronic Systems and Microprocessor Electronic Systems. His research interests are mainly related with the development of machine learning algorithms for signal and image processing. Please visit {http://www.uv.es/jordi/} for more information.
\end{IEEEbiography}

\begin{IEEEbiography}[{\IG[width=1in,height=1.25in,keepaspectratio]{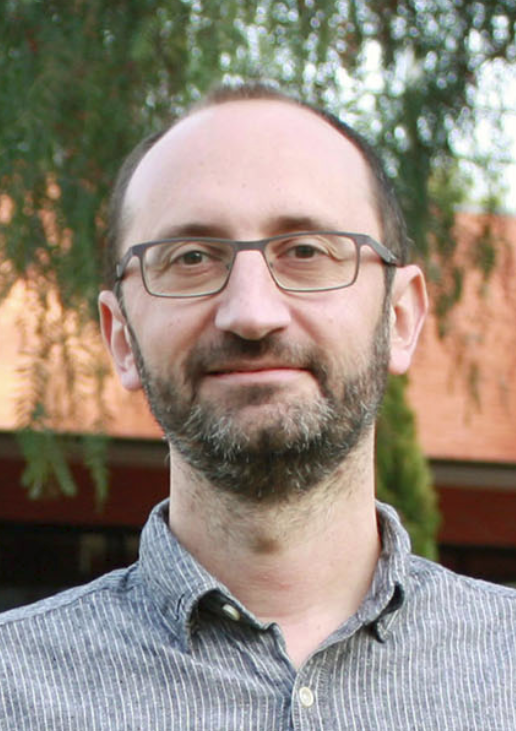}}]
{Gustau Camps-Valls} (M'04, SM'07, FM'18) received a Ph.D. degree in physics from the Universitat de Val\`encia, Spain in 2002. He is currently Full professor in electrical engineering, and Coordinator of the Image and Signal Processing Group, Universitat de Val\`encia. He is interested in the development of machine learning algorithms for geoscience and remote sensing data analysis. He entered the list of highly cited researchers by Thomson Reuters in 2011, holds an h=55, has published more than 150 journal papers, 200 conference papers, and 4 books on machine learning, remote sensing and signal processing. In 2015 he received the prestigious ERC Consolidator Grant to advance in statistical inference for Earth observation data analysis. He is Associate Editor of the ``IEEE Transactions on Signal Processing'', ``IEEE Signal Processing Letters'' and ``IEEE Geoscience and Remote Sensing Letters''. Visit http://isp.uv.es for more information.
\end{IEEEbiography}

\begin{IEEEbiography}[{\IG[width=1in,height=1.25in,keepaspectratio]{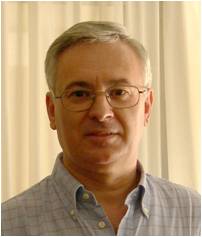}}]
{Jos\'e Moreno} is currently at the Department of Earth Physics and Thermodynamics, Faculty of Physics of the University of Valencia, Spain, as Professor of Earth Physics, teaching and working on different projects related to remote sensing and space research as responsible for the Laboratory for Earth Observation. His main work is related to the modelling and monitoring of land surface processes by using remote sensing techniques. He has been involved in many international projects and research networks, including the preparatory activities an exploitation programmes of several satellite missions (ENVISAT, CHRIS/PROBA, GMES/Sentinels, SEOSAT) and the Fluorescence Explorer (FLEX), ESA's 8th Earth Explorer mission. Dr. Moreno has served as Associate Editor for the IEEE TRANSACTIONS ON GEOSCIENCE AND REMOTE SENSING (1994-2000) and has been a member of the ESA Earth Sciences Advisory Committee (1998-2002), the Space Station Users Panel, and other international advisory committees. He is Director of the Laboratory for Earth Observation (LEO) at the Image Processing Laboratory / Scientific Park.
\end{IEEEbiography}


\end{document}